\documentclass[showpacs,prl,onecolumn,aps,superscriptaddress,preprintnumbers,nofootinbib]{revtex4}
\usepackage[T1]{fontenc}
\usepackage[latin9]{inputenc}
\usepackage{amsmath,amssymb}
\usepackage{epsfig}
\usepackage{graphicx}
\usepackage{amsmath}
\usepackage{amsfonts}
\def\slashchar#1{\setbox0=\hbox{$#1$}     		
   \dimen0=\wd0                                 	
   \setbox1=\hbox{/} \dimen1=\wd1               	
   \ifdim\dimen0>\dimen1                        	
      \rlap{\hbox to \dimen0{\hfil/\hfil}}      	
      #1                                        	
   \else                                        	
      \rlap{\hbox to \dimen1{\hfil$#1$\hfil}}   	
      /                                         	
   \fi}

\renewcommand{\vec}{\boldsymbol}
\newcommand{\beq}{\begin{equation}}
\newcommand{\eeq}{\end{equation}}
\newcommand{\bea}{\begin{eqnarray}}
\newcommand{\eea}{\end{eqnarray}}
\newcommand{\baa}{\begin{array}}
\newcommand{\eaa}{\end{array}}

\renewcommand{\vec}[1]{\boldsymbol{#1}}

\begin{document}

\title[Lev Lipatov: my  friend and renowned physicist]{Lev Lipatov: my   friend and renowned physicist\label{ch1}}
\author{Eugene ~Levin}
\email{leving@tauex.tau.ac.il, eugeny.levin@usm.cl}
\affiliation{Department of Particle Physics, School of Physics and Astronomy,
Raymond and Beverly Sackler
 Faculty of Exact Science, Tel Aviv University, Tel Aviv, 69978, Israel}\affiliation{Departamento de F\'isica, Universidad T\'ecnica Federico Santa Mar\'ia, and Centro Cient\'ifico-\\
Tecnol\'ogico de Valpara\'iso, Avda. Espa\~na 1680, Casilla 110-V, Valpara\'iso, Chile}

\date{\today}

\keywords{}\

\pacs{ 12.38.Cy, 12.38g,24.85.+p,25.30.Hm}

\begin{abstract}
These notes are written for the book
``From the past to the future: the legacy of Lev Lipatov", editors: Jochen Bartels et all, which will be published  by WS.  I tried 
  to share with you the atmosphere and the flavour of everyday life in  Gribov's theory department, where Lev matured as an independent researcher and wrote all his breakthrough papers.  \end{abstract}
\maketitle

\vspace{-0.5cm}
    \begin{figure}[h]
   \begin{center}
 \leavevmode
    \includegraphics[width=10cm]{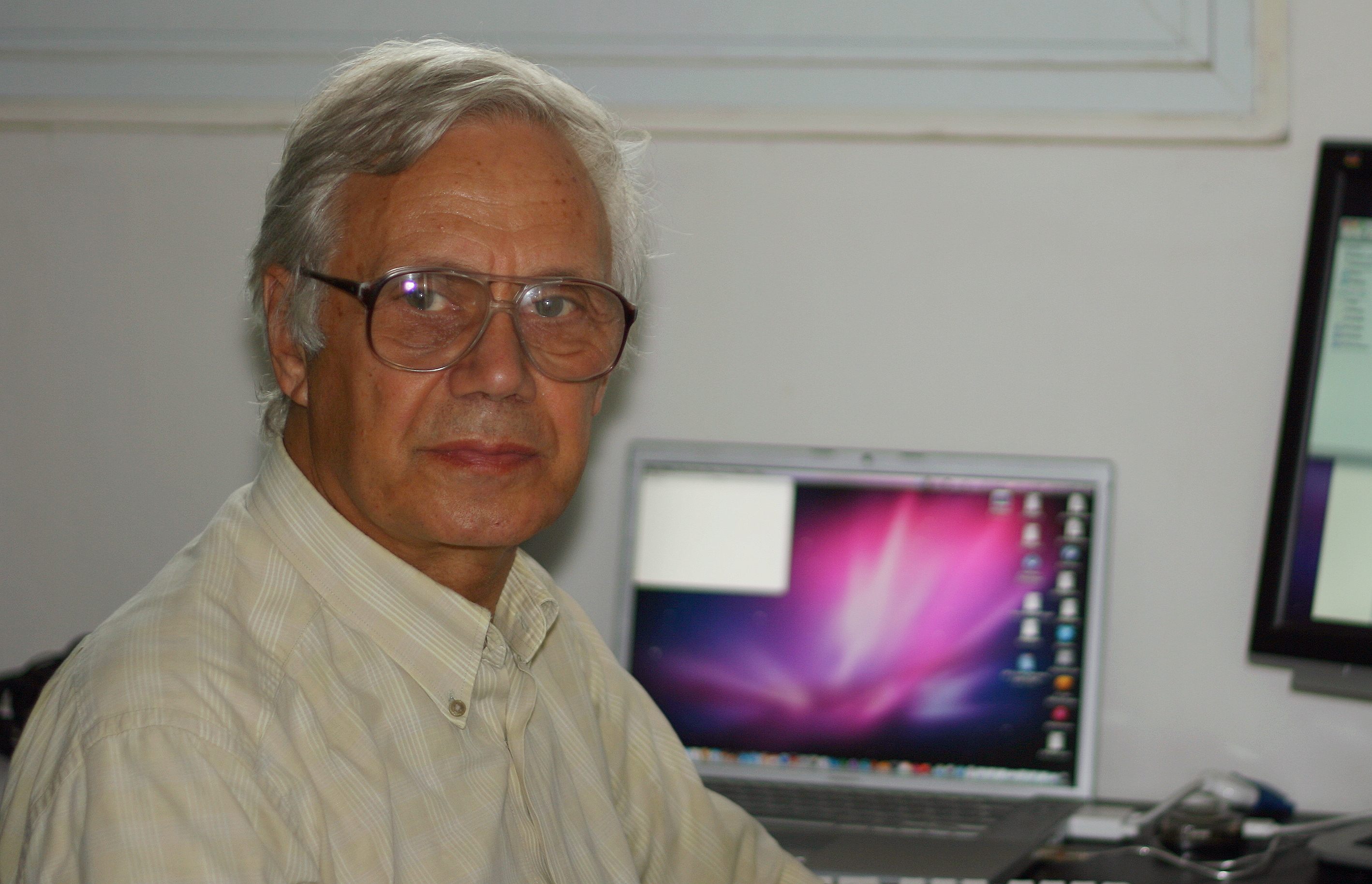}\\ 
    \end{center}
   \end{figure}
\tableofcontents
\section{Introduction}\label{sec1.1}
\begin{figure}[h]
   \begin{center}
 \leavevmode
    \includegraphics[width=10cm]{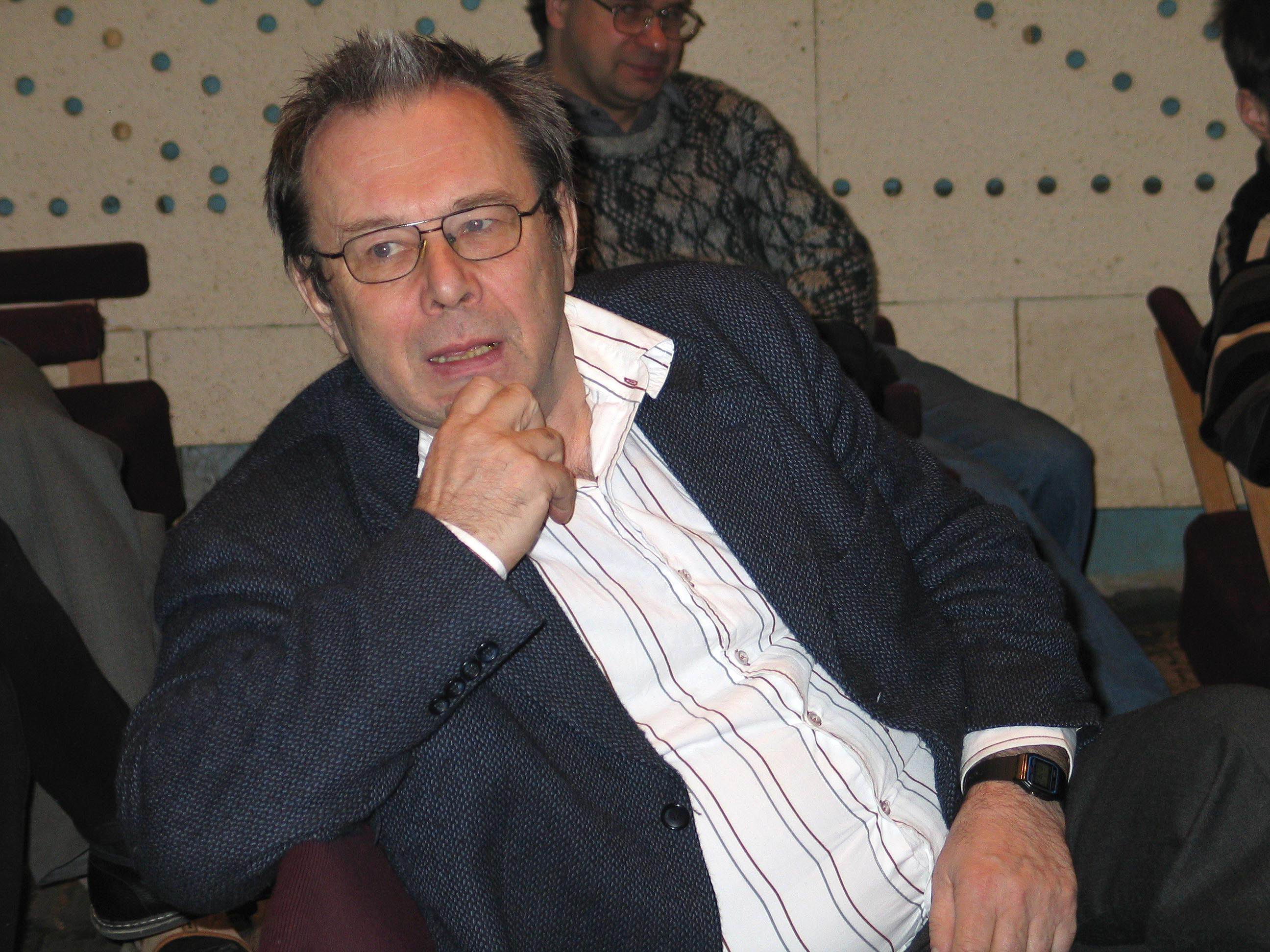}\\ 
    \end{center}
    \caption{ Quam bene vivas refert non quam diu (It is how well you live that matters, not how long)}
    \label{lip1}
   \end{figure}

 Prof. Lev Lipatov (see Fig.\ref{lip1}) died on September 4 2017 during a conference in Dubna.  Physicists all around the globe, who are doing  high energy physics, and, especially, Quantum Chromodynamics,    know or feel that this area of physics would be quite different, without his contributions,  and the future progress will be slower without his ability to solve difficult problems. I believe, that this book will show  you,  what we lost as far as physics is concerned. The goal of my presentation is quite different: I wish to share with you the atmosphere and the flavour of everyday life in  Gribov's theory department, where Lev matured as an independent researcher and wrote all his breakthrough papers.  Actually, his entire life was in this department, where he started as Gribov's Ph.D. student, and finished as the head of the department.  If you wish, my paper deciphers the main credo, which we were taught by Prof. Gribov and by the physicists of his generation: {\it ``Physics  first"}\footnote{It should be pronounced as `ladies first' in a chivalrous code. Meaning is obvious: we think first about physics. It is so natural as to let a lady to go first.}
 
\section{The beginning}\label{sec1.2}
\begin{figure}[!tbp]
  \centering
  \begin{minipage}[b]{0.4\textwidth}
    \includegraphics[width=0.8\textwidth]{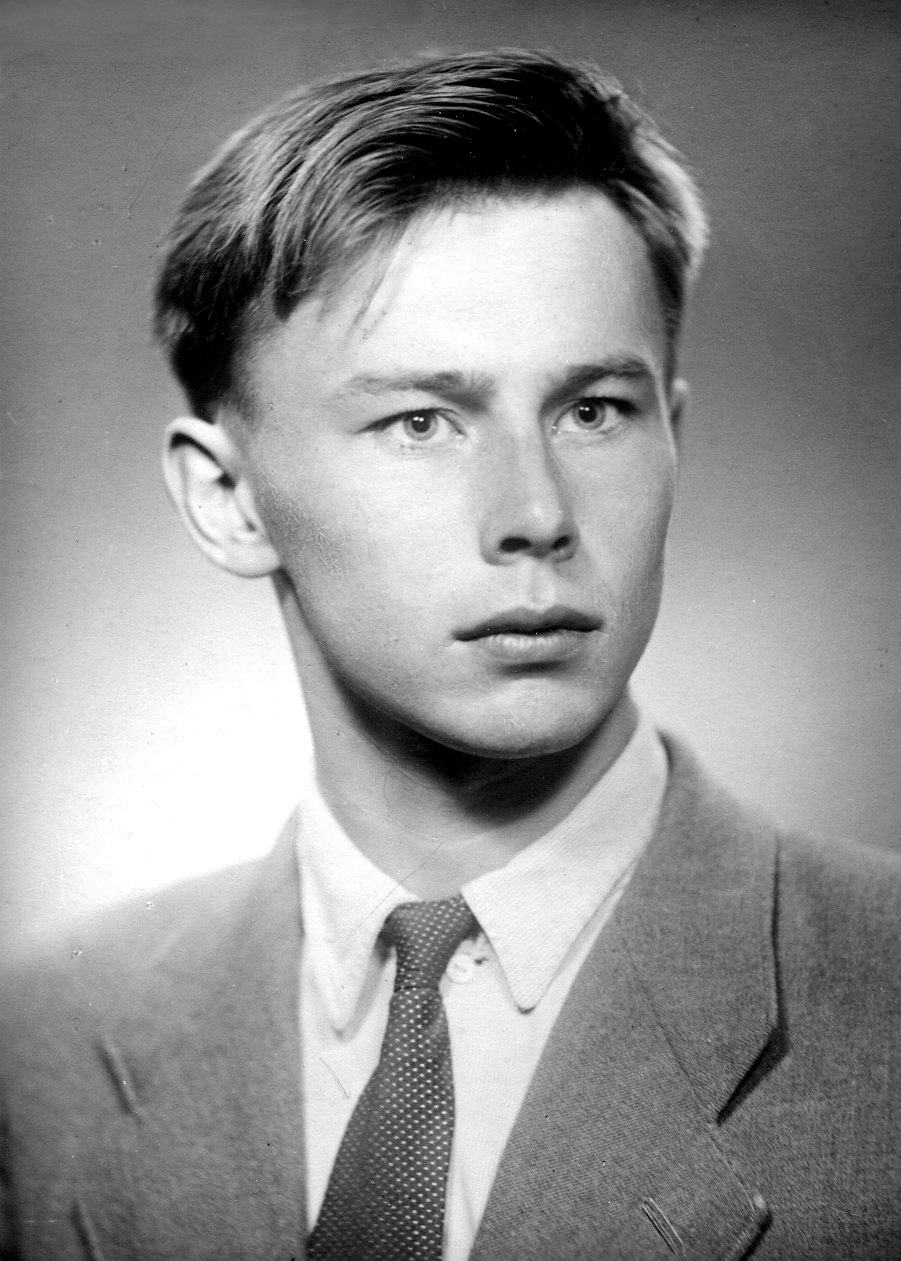}
    \end{minipage}
      \begin{minipage}[b]{0.4\textwidth}    \caption{A man comes into this world with clenched fists and a cry: he seems to be saying the whole world will be mine: Lev Lipatov, student of the physics department of S.Petersburg University.}
      \label{lip2}
  \end{minipage}
  \end{figure}

Let me start from the beginning. The first time I met Lev was  in July 1957. It was a  rare  sunny day in Sankt Petersburg (Leningrad at that time) in the country, which does not exist: in the Soviet Union. We did not pay too much attention to the beauty of the day, since we had to write the entry exam in physics to  the  Physics Department of S.Peterburg University. Nowadays, it is difficult to believe, that at that time physics was so popular in Russia, that every mother wished that her daughter would marry  a physicist. The competition was very strong: ten people for one place. Therefore, the five problems, that we had to solve, were  rather difficult, especially one. When I finished, I went out asking around about the answer to the most difficult problem. The first three  gave me an  answer, which was different from mine. The fourth was Lev, who had the same answer as I had. In three days, we found ourselves among the people, who passed the exam( see Fig.\ref{lip2}). 

  \hfill
  \begin{figure}[!tbp]
  \centering
  \begin{minipage}[b]{0.7\textwidth}
    \includegraphics[width=0.9\textwidth]{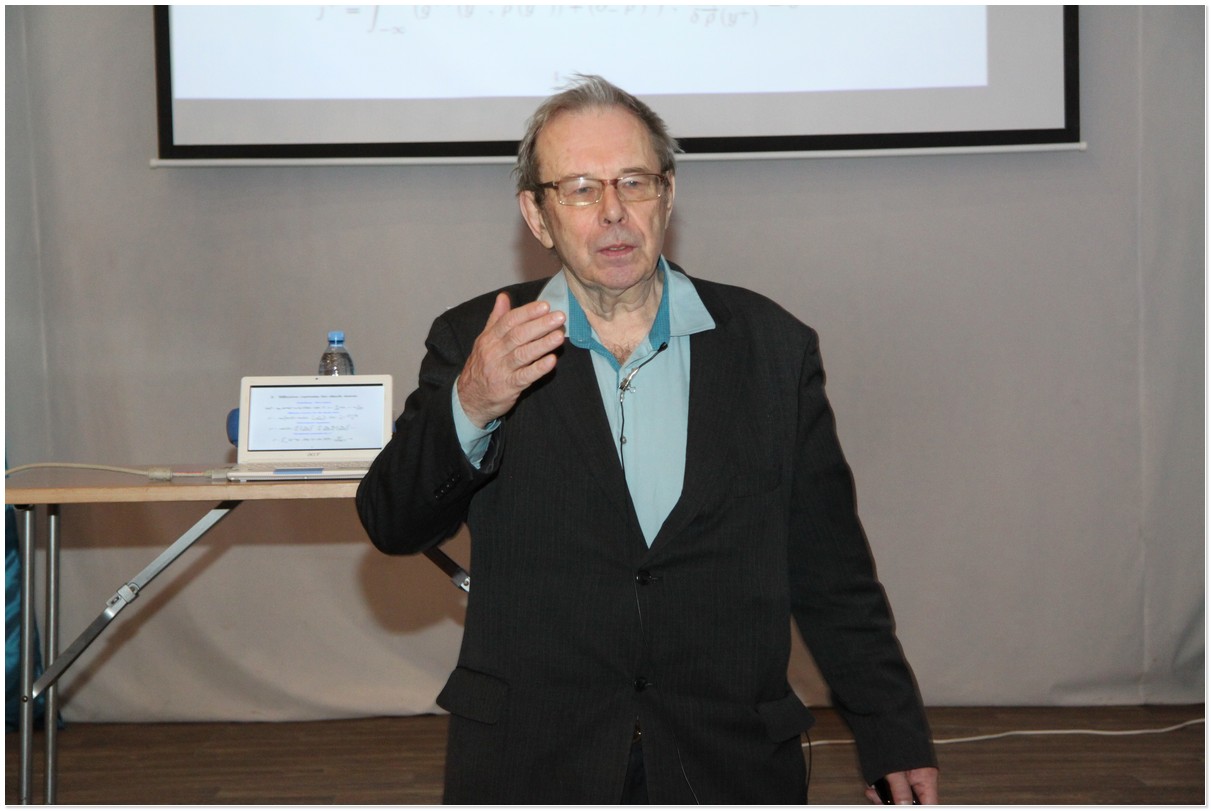}    \end{minipage}
      \begin{minipage}[b]{0.2\textwidth}    \caption{And he leaves with open hands: as if he says, I don't take anything with me: one of the last pictures of Lev that I have.}
      \label{lip3}
  \end{minipage}
  \end{figure}

 Starting from that day, we were in touch on an  everyday basis for the rest of our lives. We went  together to the same group for the general course of physics, together decided to do theory of elementary particles, and went to the theory group at the university. We created a  net of  common friends in the university  and found  our  wives among them.
 
\begin{figure}[h]
   \begin{center}
 \leavevmode
    \includegraphics[width=10cm]{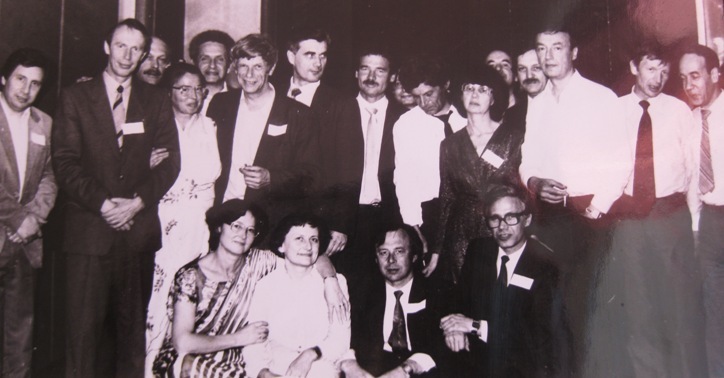}\\ 
    \end{center}
    \caption{We and our girls: theory group of the physics department of S.Petersburg university, 25 years after graduation. Lev and I are in the first row.}
    \label{lip4}
   \end{figure}

  Our friends (see Fig \ref{lip4}), they deserve 
more than several lines in the paper. They shared with  us  all our  ups and downs, they supported us without questioning whether our behavior was correct or not.  I know, that Lev was proud, that we have not lost even one friend during 60 years after the university. Even now, when I am writing this paper, sitting at home in Israel, I hear  my wife  discussing the important matter like children and grand children with her friends in S. Petersburg, over  Skype.

During the university years the mathematical capability of Lev was fully recognized  by us.
Actually, he wanted to get  a major in the mathematical physics. Fortunately for the theoretical physics,  he changed his plans and decided to do high energy physics. I believe, that it happened
because he realized that the high energy physics is full of difficult  unsolved problems. One episode was very instructive, which shows that we were  very well motivated kids. Prof. Yu.V. Novozhilov, who taught the course in quantum field theory, chose to follow the variational approach of J. Schwinger  to quantum mechanics and quantum field theory. Frankly speaking,  it was very difficult for us and,  when we found the book of Bethe, Hoffman and Schweber: ``Mesons and Fields", that  had been recently translated to Russian, we decided to ask the young star of the theoretical department Prof. Mikhail Braun to organize an informal seminar in which we read and discuss this book. I remember, that three of us, Alexander Vassiliev, Lev Lipatov and I, were brave enough to go to Prof. Braun and ask him to help us. He agreed and most of my friends, including Lev and myself, consider him as our first teacher. This paper is a good opportunity for me to express in public, how we were and are thankful to him. I believe, that after this event Lev changed his mind and joined us in our attempts to learn quantum field theory.

Impressed by the personality of Prof. Braun, we decided to write the diploma paper with him. This was a mistake. Neither Lev nor I,  remember the subject of this paper, which we did together. It was never published.  The only result from this work was, that we learned some basics of the  Feyman diagrams approach,  and found out,  that the only place in S.Petersburg, where the high energy physics was on a good international level, was the Gribov department in Ioffe Physico-Technical Institute. After getting our masters degrees, we were allocated to the research institutes in S. Petersburg, in which we were involved in quite different physics. However, we were shy but stubborn kids, and during  1963  we  started to pass through  the so called Landau exams in theoretical physics. Our  goal was to get a position in the Gribov department, but it was only possible after passing through the set of exams. For us, since we were fresh from the university, Gribov cut down a bit the set of exams and we needed to pass: classical mechanics, classical electrodynamics and quantum mechanics, based on Landau and Lifshitz book, and quantum electrodynamics from the book of Akhiezer and Berestetsky.  It was a hard time for our young families, but working at night, after the day in our institutes we passed  all exams, solving  difficult problems.  Prof. Gribov took  us as his the only Ph.D. students. Gribov's third Ph.D. student, L. Frankfurt, also passed the exam procedure. This was the end of the beginning for the  three of us.

\section{Gribov's department and high energy physics seminar.}\label{sec1.3}

Let me start with two citations. The first one is from Wikipedia: ``
Vladimir Naumovich Gribov was a prominent Russian theoretical physicist, who worked on high-energy physics, quantum field theory and the Regge theory of the strong interactions.
His best known contributions are the Pomeron, the DGLAP equations, and the Gribov copies. "
The second is  from the paper of  Dokshitzer and Kharzeev \cite{DOKH}:  ``
Suffice it to mention that his name is attached to many  key notions of the theory arsenal: Gribov-Froissart projection and the Gribov vacuum pole (Pomeron),
Gribov factorization, Reggeon Calculus and Reggeon Field Theory (RFT), Gribov diffusion, the AGK (Abramovsky-Gribov-Kancheli) cutting rules, the Gribov bremsstrahlung theorem, Gribov-Glauber theory of relativistic multiple scattering, Gribov-Lipatov evolution equations, Gribov copies and the horizon, etc ."
The dry language of these citations shows,  that we were lucky to get  as our supervisor an  outstanding physicist, but they do not describe the Gribov personality, his drive for physics, his intuition and ability to create a picture for  complicated physics problems. This is not a paper about Gribov, but his influence on Lipatov and us is difficult to overestimate. Gribov, whom we know and respect, is described by Dokshitzer in Ref.\cite{DOK} and I can only certify, that he gave a portrait of our teacher, which is close to reality.

We came when Gribov has published his famous papers on the Pomeron and Gribov - Froissart projection, but other items, listed above,  were done in our time  and with our participation.

 However, our first problem  was to survive, when we came to the department after passing the  Landau exams.  Prof. Gribov summoned the  three of us and gave the following instructions. You have two duties: to come and participate in any seminar of the department; and to answer all questions that an experimentalist would ask you. You are free to do whatever you want, but do not think, that I will find  a problem for you. You need to find it by yourselves. However, you have a privilege: you can ask me any questions and I will answer.   Therefore, we understood that 
 Prof. Gribov used the principle  of deep water in his pedagogical approach. You know, that there is a method to teach kids to swim: to hurl then in the deep water. They say that some of them will start to swim.

    \begin{figure}[h]
   \begin{center}
 \leavevmode
    \includegraphics[width=7cm]{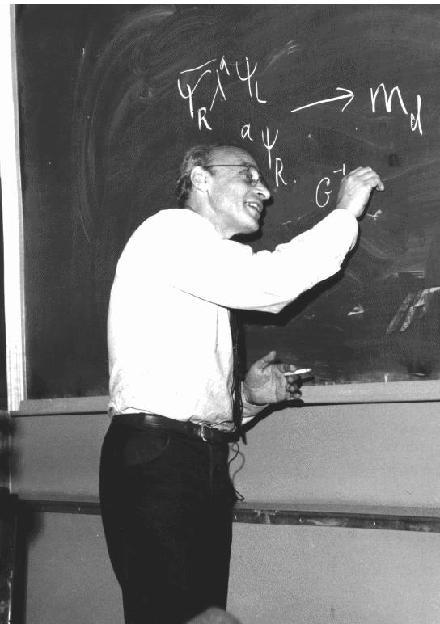}\\ 
    \end{center}
    \caption{Magister dixit: Prof. B.H. Gribov in our common room is telling us,  what he understood . The best picture that I found which reflects his personality}
    \label{lip5}
   \end{figure}

We survived,  but many did not. Lev and I sweared,  that if we would survive and will have our own students, they will never suffered as we  did.  My former students, S. Bondarenko and  A. Prygarin, and the young researcher in UTFSM, M. Siddikov, with whom Lev wrote his last papers, have no idea,  that the  patient and attentive attitude of Lev to them,  was formed in  distant 1964 due to  our experience.

In 1963,  when we joined the department, Gribov's department  in Physico-Technical Institute  was a unique    phenomenon having about 15-20 men, who were young (all, except Ilia Mironovich Shmushkevich, were younger or the same age as Gribov, who was 33, when we came), ambitious, creative and well educated.
All of them were doing the same physics as Gribov. We actually worked at home having two visiting days on Mondays and Thursdays. On Monday was the Gribov seminar, which I will discuss below, but on Thursday the informal discussions went from the morning till 6 p.m.  We had only one common room\footnote{Actually, this fact: one room, was the official argument,for  why we had only two visiting days.} in the institute and the typical  scene on Thursday was Gribov relating, what he did or thought during the days from the previous meeting on  Monday (see Fig.\ref{lip5}). We all were listening and trying to understand and argue against. For us, youngsters, it was difficult to digest, what he was talking about, and even much more difficult to take parts in the discussions. 

Fortunately, there were three of us and each of us worked hard. Our internal chatting with attempts to understand the new ideas and accumulate them, was the main factor, that helped us a lot.  The second factor was that Gribov was not strict with his promise not  to meddle in our search for the subject: Frankfurt and I  got the idea from him to look for hadron scattering in the additive quark model using the analogy with deuteron-deuteron collisions; Lev was taken by him to the team  for reconsidering double log approximation in quantum electrodynamics.

This department had a regular seminar, which was  unique as the department itself. After leaving Russia in 1990, I have travelled a lot, but have not seen anything, which is close to Gribov seminar.  The goal of the seminar was not to learn, what the invited speaker did, but to understand the physics  problem and find the solution to this problem. Sometimes with the help of a speaker, sometimes without his help. It was assumed,  that: (i) the speaker is a good physicist, since he was invited to give a talk;(ii) he wants to convince us, that he made a breakthrough, but with  high probability, he is wrong; (iii) the speaker is responsible for everything, that he used and should prove every point of his talk\footnote{If somebody would ask him to prove the Pythagoras theorem, the speaker should prove it.};(iv) the speaker must answer any questions, related to his talk.  The big shots and the young students are equal in giving the talk and in asking the questions.  Your past merits cannot save you from  deserved criticism. The seminar starts at 10 a.m. on Monday. The speaker has 5 min. of tolerance, during which he should not be interrupted.  The duration of the seminars  were not limited. I did not remember a seminarthat be finished before 2 p.m. but very often after lunch break it continued to 6 p.m.. I remember several cases, when we gathered together on the next day to continue the discussions.

The typical talk at the seminar was as follows. At 10 a.m. Gribov introduces the speaker. Five minutes the speaker talks. After that Gribov jumps up saying first, the speaker is completely wrong,  and  second, that correct answer should be like this. The description of the answer followed, taking 30 mins. After such beginning the weak speakers started to be nervous, but the ones, who trust themselves, answered , that you, Prof, Gribov,  are completely wrong and  that you have not taken into account this and that. Other people enter the discussion. Sometimes the discussion turns out to be so hot, that everybody forgets about the speaker. Normally, after two hours of discussions, the common opinion was to let the speaker talk as he wants, but to emphasize the points that have been raised in the discussions.

It was a painful experience for anybody,  who talks at this seminar,  an ordeal not only for your knowledge and ability in physics, but for your character. We know this from our own experience, since everyone of us had to report on his paper at the seminar before sending it for  publication. However, after the talk, when the dust had settled, 
you realized that actually    the friendly fight, which you had at the seminar,  was very useful for you: you were able to look at the problem, that you solved in the paper, at different angle, which would help you both in writing the paper and in the future work.

After 1985 we started to go abroad and many of us found jobs in the West. We found ourselves quite capable to discuss and fight for our ideas with the experts, but were very badly prepared to give a talk for a more general public: for the physicists, who were doing the high energy physics, but whose interests were rather far away of ours.  Prof. Lev Lipatov suffered from this problem to the end of his days. His talks were very useful and enjoyable for a couple of dozen of experts ( most of them are the authors of this book), but were rather 
a black box for the others. He actually recognized this problem and worked hard to solve it.  Our bad luck, that he left us before solving it.

 My last comment in this section is about our  second  duty: to answer {\bf all} questions of an experimentalist. The origin of this demand lies in the strong belief, that physics is the experimental science.  Nature is so beautiful, that we do not need a substitute: the formalized theory, let say the string theory, as an example. Our teachers demanded of us, that we know the mathematics, which is a tool for solving the problem, but the main task of  physicists is  correctly  formulate the physics problem, which will be beautiful and solvable only because it is  part of  nature. Lev Lipatov spent the most of his life solving  theoretical problems, but he evaluated very high his attempt with D. Ross and H. Kowalski (see the paper in this book) to describe the HERA data.  He told me, that finally, he feels himself as a normal physicist, who is  investigating  nature, not only  writing  abstract formulae.

 \section{ 1966 - 1974: Double logs in QED (The log academy and its president)}
 
 As I have mentioned, Gribov added Lev to his team, which reconsidered and developed the double log approximation for QED\cite{DLA1,DLA2, DLA21,DLA3,DLA4}. It was a lucky chance for Lev. His mathematical talent, which was known to his friends, found  a fertile ground for improvement and development. Frankly speaking, at that time we looked at him  a bit from above. Indeed, we were involved in such important issues as additive quark model, finite  energy sum rules, duality models, Reggeon calculus, parton and multiperipheral models and so on. Field theory was considered  as the past. Ten years later the situation reversed itself.
  
  We all started to re-learn the field theory with the advent of QCD. Lev Lipatov was  better prepared  than everybody else, even better than Gribov himself, as you will see from the next section. I cannot express in a proper form how much I, personally, learn from Lev about QCD and secrets of the  perturbative approach to it. Lev's contributions to team work was recognized and he defended his first dissertation (candidate of science) in 1968 on the basis of Refs.\cite{DLA3,DLA4,DLA5,DLA6}\footnote{I believe, that these papers give a good insight in comparison of  Russian Candidate of Science degree with the Ph.D. one. I think, that in the best Russian institutions the C.Sc. degree was  more or less the same as Ph.D.}.  For a long time Lev continue to work in calculating different processes in QED (see Refs.\cite{DLA7,DLA8,DLA9,DLA10,DLA11,DLA12,DLA13,DLA14,DLA15,DLA16,DLA17,DLA18,DLA19}). I put the titles of the papers, that you can see the diversity of Lev's  interests.  Calculating these QED processes Lev became     a great expert in finding and summing  log contributions in  perturbative QED Feynman diagrams. 
  
  The double log approach consists of two ingredients. First, in each Feyman diagram of  order  $\alpha^n_{e.m.}$ in QED  the contribution, which is proportional to $( \alpha_{e.m} \ln s \ln t )^n$, has to be calculated.  Second, all these contributions have to be summed as an analytical function, mostly, writing the equation which sums all these contributions. In brief
  \begin{equation}  
  \mbox{DLA}: A( s, t) \,\,=\,\,\sum_n \,C_n\, (  \alpha_{e.m} \ln s \ln t )^n
    \end{equation}
for 
      $\alpha_{e.m} \ln s \,\ll \,1$    and  $ \alpha_{e.m} \ln t\, \ll\,1$ while  $  \alpha_{e.m} \ln s \ln t\,\sim\,1$.
      
   The technique, which is used, includes the Sudakov decomposition: 
        \begin{equation}  
 k_\mu\,\,=\,\,\alpha p_{1,\mu} \,\,+\,\,\beta\,p_{2,\mu}\,\,+\,\,\vec{k}_{\bot}
    \end{equation} 
    where $p_1$ and $p_2$ are the momenta of colliding particles and $k_{\bot,\mu} p_{1,\mu}=0$ as well as $k_{\bot,\mu} p_{2,\mu}=0$. Lev was not only very skillful in finding log integrations in $\alpha$ , $\beta$  and $k_\bot$,  but he 
    introduced  the analyticity and unitarity, which were often discussed in our department,
    in  perturbative calculations. It was new at that time.
             
   We recognize his superiority and joked, that he creates a log academy, where he is the president and E. Kuraev and V. Fadin are the academicians. Humor is humor,  but I was proud, when in 1982 they incorporated me as the corresponding member of this academy.

   Doing his QED papers, Lev found  and created his own team: E. Kuraev (from Khar'kov) and V. Fadin(from Novosibirsk). Below you will see how this team  will work.  Edik Kuraev has left us and I wish to remind, that he came to our department from Khar'kov and quickly understood, that this is his chance to learn physics. He worked as a night guard in S. Petersburg to earn money and stayed with us for long time. As you can see from the list of references,  Lev did a lot of work with him on QED. He also initiated the Gribov's lecture on strong interactions, which he printed as  a Khar'kov preprint. He is the letter $K$ in the celebrated BFKL Pomeron, which we will discuss below. He was not well known mostly because he could not speak English, but he was the key member of the Lipatov's team. It would make me happy, if you consider these several words as a tribute to him.

 \section{1971-1974: DGLAP evolution equation}
 
 In 1969-1970 four papers of    S.~D.~Drell, D.~J.~Levy and T.~M.~Yan 
 were published in Phys. Rev.\cite{DRELL}, in which they demonstrated, that the main features of the parton model  can be derived from quantum field theory assuming, that all transverse momenta are cutoff at some value, which does not depend on energy. These papers impressed Prof. Gribov,  and he started to discuss, what happens with the deep inelastic processes , if we do not make a momentum cutoff. Every Monday and Thursday we heard about his ideas on this subject and finally, it has been announced a seminar where he would present his findings.
 
 The seminar proceeded in usual fashion. Gribov explained, why he thinks, that the correct integration over transverse momenta will result in log contributions ($\ln Q^2$, where $Q^2$ is the photon virtuality in inclusive $\gamma^* p $ scattering).   He proposed to calculate the deep inelastic structure function as\footnote{The modern notations are used in 
 Eq. 1.3: $ x \,=\,Q^2/s$ where $Q$ is the photon virtuality, $s$ is the energy squared. $\alpha_s$ is QCD coupling which in the paper of Gribov and Lipatov is denoted by $g^2$.}
 \begin{equation}\label{2}
 F( x, Q^2)\,\,=\,\,\sum_n C_n(x) ( \alpha_S \,\ln Q^2)^n
 \end{equation}
 assuming $\alpha_S \,\ln Q^2\, \sim \,1$ while $\alpha_s \,\ll\,1$ and $\alpha_s \ln(1/x)\,\ll\,1$.

 He started to tell us, how to sum all these logs,
 but in five minutes Lev Lipatov  raised his hand and very politely told: ``Vladimir Naumovich, you are wrong with your procedure to find and sum the logs"; and in ten minutes he explains why and what we need to do to find the sum. When he finished everyone understood two things: first, Prof. Gribov is, indeed, wrong; and second, Lev Lipatov is the one who will be the head of this department after Gribov. These ten minutes changed  the standing of Lev from the expert in QED calculations, which were not the 
 primary interest of the group, to the main expert in quantum field theory.
 
 Frankly speaking Gribov tried to answer Lev and defend his point of view. It lasted about 40 min. Finally, answering one of the question, he gave up and    said: `` Lev, you are the part of all this, help me to answer."
 
 It goes without saying,  that this seminar was the shortest that I remember.
 In one month time, a new seminar was held,  in which Lev Lipatov was the speaker. It was also one of shortest seminars, since Gribov was very quite and supportive. Lev told us how to sum logs of transverse momenta
 and derived the equation for the deep inelastic function from which we see the violation of the Bjorken scaling behaviour due to these log contributions\cite{DGLAP,DGLAP1}. The estimates were made  in vector and pseudo-scalar meson theories, which were on the market at that time, but all essential elements: gauge invariance, violation of the Bjorken scaling
 and so on, were found. Frankly speaking, when QCD appeared the only  item  that we needed, was to calculate gluon $\to$ gluon + gluon splitting function. However, both Gribov and Lev were very busy with different problems: Gribov, was doing his copies, which is really more important, than the correct evolution equation. Lev answered  my suggestion, that he can calculate the gluon splitting and write the QCD equation in two days, saying that he does not have a good student for this simple job  but, ``I am close to understanding what is the Pomeron in gauge theories".  Therefore, the correct evolution equations were waiting for a good student. Finally, Alexej Anselm, gave this problem to his good student: Yura Dokshitzer, who wrote the QCD evolution equation\cite{DOKSH} but in 1977,  it was done independently  by Altarelli and Parisi\cite{AP}.

 I would like to quote here the citation from Wikipedia about QCD evolution equations: ``DGLAP (Dokshitzer-Gribov-Lipatov-Altarelli-Parisi) are the authors who first wrote the QCD evolution equation of the same name. DGLAP was first published in the western world by Altarelli and Parisi in 1977 \cite{AP} hence DGLAP and its specializations are sometimes still called Altarelli-Parisi equations. Only later did it become known that an equivalent formula had been published in Russia by Dokshitzer also in 1977\cite{DOKSH} and Gribov and Lipatov already in 1972\cite{DGLAP,DGLAP1}".

I do not think that this citation correctly reflects the situation. Certainly, experts knew about the Gribov-Lipatov approach before 1977. It is enough to mention that one of the principle papers, that sets the formalism of the operator product expansion for the deep inelastic function (see   N.~H.~Christ, B.~Hasslacher and A.~H.~Mueller, Ref.\cite{MUDIS}) is based on the Gribov-Lipatov equations. I think, that situation is different, people in  the West did not know the next paper of Lipatov\cite{LIPDIS}, in which he explained how  the partons  appear in the gauge theory.  About partons they learned from Altarelli and Parisi,  and this fact explains, why Altarelli-Parisi paper was so popular.

In 1974 he was the first among  us to obtain the degree of Doctor of Science\footnote{The second degree in Russia, which is similar to habilitation in Germany. You cannot get a full professor position without this degree. Lipatov, once more, gives a clear example what it meant to be Doctor of Science in Russia at that time.} for these equations.

In 2015  High Energy and Particle Physics Prize, for an outstanding contribution to High Energy Physics, is awarded to
James D. Bjorken   for his prediction of scaling behaviour in the structure of the proton that led to a new understanding of the strong interaction, and to
Guido Altarelli, Yuri L. Dokshitzer, Lev Lipatov, and Giorgio Parisi  for developing a probabilistic field theory framework for the dynamics of quarks and gluons, enabling a quantitative understanding of high-energy collisions involving hadrons. 

I believe, that the above citation  nicely summarizes the achievement of Lev and 
the importance of his work for the high energy community. It goes without saying, that Lev Lipatov got  international recognition after this work, and his papers jumped over the iron curtain barriers and became known in the West. However, as I have mentioned, his main contribution to DIS\cite{LIPDIS}  remained unknown till 80's , when we started to travel abroad.
    \section{ 1975-1990:   Pomeron in QCD (BFKL Pomeron)} 
The Pomeron structure was   Lev's entire life's work.  The high energy asymptotic behaviour of the scattering amplitude,  which in the Regge approach was related to the exchange the Pomeron, was discussed in the Gribov department every visiting day and was considered as the key principle question of the high energy physics. Thinking about this back in time, frankly speaking, I think it was, to a large extent, the relic of the non field theory(FT)  approach, which was the standard in 60's.  The idea how to approach a high energy amplitude in the FT
 was suggested in Ref.\cite{GLF} by Gribov, Lipatov and Frolov in QED, but the following simple equation clarified this approach in QCD:    
       
\begin{equation}  \label{LLSA}    
{\rm L L(s)A} :~~~~~~A( x,t) \,\,\,=\,\,\,\sum C_n(t) \Big( \bar{\alpha}_S \ln (s)\Big)^n; ~~ \bar{\alpha}_S\,\ln (t)\,\ll\,1;~~ \bar{\alpha}_S \ll 1;
\end{equation}
where $t$ is the momentum transferred squared.

\begin{figure}[h]
   \begin{center}
 \leavevmode
    \includegraphics[width=12cm]{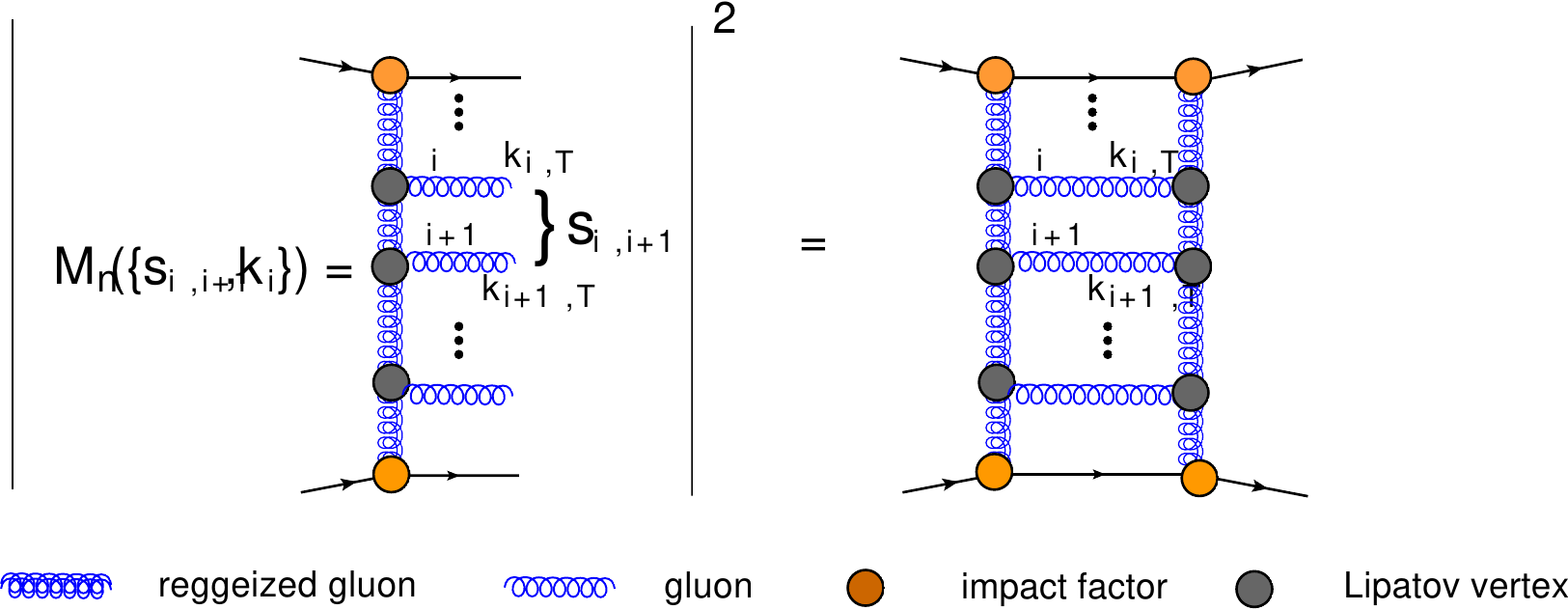}\\ 
    \end{center}
    \caption{BFKL Pomeron and its main ingredients}
    \label{bfkl}
   \end{figure}

Over four years Fadin, Kuraev and Lipatov \footnote{In our department the names of authors follow the alphabeting order. Since the main papers  were  written in Russian, one needs to know  that  the letter F is one of the last letter  in the Russian alphabet, while in English it is one of the first.  I follow  the English order, which is commonly accepted for  the BFKL Pomeron.}  developed the  leading log(s) approximation for non-abelian gauge theories from calculating the  diagrams in the first orders of $\alpha^n_S$ to the solution of the general equation\cite{FKL1,LIPBFKL,FKL2,FKL3}. I believe that the numerous features of this equation will
be the subject of the most papers in this book. Therefore, I am  not going to discuss them.
I wish only to give you a flavour of the hot discussion in our department on this equation.
Nowadays, most physicists think about the BFKL  equation as the evolution of the DIS structure function in the region of small $x$. Lev, when he discussed this equation, did not see this application. For him it was an attempt to solve the Pomeron problem and to answer the question, what is the structure of the Pomeron in  the gauge field theories. Fig.\ref{bfkl} shows the main ingredients of this structure. The left figure shows  that the FKL  production of $n$  gluons at high energy can be described as the ladder diagram, and has the following expression:

\begin{equation} \label{GPROD}
M_{\rm 2\, \to \,n }\,=\,\prod_{i =0}^{n - 2}\Big(\frac{s_{i-1,i}}{s_0}\Big)^{\alpha_G(q_i)} \Gamma_L(k_{i,T})
\end{equation}
where $s_{-1,0} = s$ and $ q_i$ are the transverse momenta of the vertical reggeized gluons. $\alpha_G( q_i)$ is the spin of these gluons. $\Gamma_L$ is the Lipatov vertex which  differs from the triple gluon vertex in the QCD Lagrangian. The contribution to the scattering amplitude is equal to
\begin{equation} \label{AMP}
\mbox{Im} A(s, t)\,=\,\sum_{n=2}^\infty |M_{\rm \,2 \,\to \,n }|^2
\end{equation}

These two equations show that FKL gave the $s$-channel picture for the Pomeron exchange and, actually, stated that the maximum contribution comes from the gluon production in the typical parton kinematic at high energies. The first surprise is that the same structure satisfies the $t$-channel unitarity. In the framework of Regge approach, we believed that the $t$-channel unitarity and the crossing symmetry are more important than the $s$ -channel unitarity, which  will come automatically, when we will build the theory, which holds unitarity in $t$-channel. Why surprise? Because it turns out that due to the exchange of a gluon, whose spin depends on it's transverse momentum (reggeized gluon), the  emitted gluons do not follow the Gribov diffusion and do not lead to the amplitude which depends on $t$. In principle, we knew, that there is such a general possibility, that the asymptotic behaviour of the amplitude will be determined by the standing singularities, but nobody expected that QCD, the long awaited microscopic theory, will produce  the result, which will be  opposite to what we  have believed  during at least ten years before its advent.

The  reggeization of gluon, together with the structure of the Lipatov vertex lead to a different kind of the diffusion process: diffusion in logs of transverse momentum. It means that in the process of evolution, the transverse momenta of the produced gluons acquire a  larger as well as  smaller transverse momenta. Therefore, from the early days  of the BFKL Pomeron we knew that at high energy we have to understand what happens to the QCD Pomeron: calculated in LL(s)A  of perturbative QCD it generates gluons with small transverse momenta, which have to be treated in non-perturbative QCD. If you take into account, that the intercept of the BFKL Pomeron turns out to be larger than 1, in the contradiction to the Froissart theorem, I hope that you understand how important  the BFKL Pomeron was and is  in high energy physics. 

Now I need to clarify the letter B in the BFKL Pomeron. As you can see from Ref.\cite{BALI}
this letter stands for Balitsky, who was the lucky guy. Lev did not want to repeat his mistake, when he refused to develop Gribov-Lipatov equation to the case of QCD. Fortunately, he had a good student, Ian Balitsky, and he gave him a problem: to re-write FKL equations ,
which they derived in Yang-Mills  SU(2) theory,
for  a more trivial case of gluons with zero mass. In this paper, they not only re-write the equation for the only interesting case, but found the exact solution for the total cross section.

Lev considered the reggeization of gluon in QCD is a most important result. He thought that the major theoretical problem was to build the effective theory for QCD at high energies, which will include the new degrees of freedom:  the reggeized gluons. He suggested the effective action (see Ref.\cite{LIPEA}), which he hoped, will solve all problems and will give the effective theory for high energy QCD. He was so firm in his belief, that he thought  that  CGC and the colour dipole approaches  are just wrong, because in the framework of these approaches the new phenomena of the gluon reggeization appears as a trivial property of the normalization of the  dipole wave function. Frankly speaking, we only once discuss this problem, when he learned, that I like both the CGC-dipole approach and the fact, that the reggeization looks so simple. I feel, that it is good, that I am in Israel not in Russia,  for keeping our relationship  on a positive level. I was really afraid that our long term friendship will be broken. All my arguments, that simplification is the main driving force of physics and that he was in the beginning of the coordinate approach, which seriously clarifies the BFKL Pomeron structure, were in vain. After Yura Kovchegov and I  wrote the book \cite{KOLEB} on high energy QCD, he was very unhappy. I was able to placate him saying, that we presented his point of view and also, he has a book, where he argued his point of view in full \cite{LIPB}. He agreed saying, that he wrote a book not for kindergarten kids.  However, I must admit,  I have not found courage to tell him, that I wrote a paper, which shows how  trivial the reggeization  is in CGC approach\cite{KLREG}.

 The twist of fate is indeed the fact that Lev first wrote the BFKL Pomeron Green function in the coordinate representation and found the solution to the BFKL equation in the general case\cite{LIPSOL}. Writing this note I looked and found  that this paper has less than a  thousand citations. I was upset, since this paper is the most essential paper in the whole business of the BFKL equations.

  \section{Lipatnik and the begining of Lipatov school}
   As I have mentioned, our duty was to answer all questions of the experimentalists. However, as it is well known, experimentalists have so many problems, besides physics, that they very often have no time to think about physics. For physicists at the universities this problem  shows up less, since they have a common seminar, where both theoreticians and experimentalists discuss their results and related physics. In the research institute, as it was in the PNPI, the seminars were  separate, more than this before 1972 the theoretical high energy physics was concentrated in Ioffe Physical-Technical Institute in S.Petersburg (Leningrad at that time) while the experimental laboratories were placed in the small town of Gatchina which is 44km to the south of the city. After 1972 the separate institute: Petersburg Nuclear Physics Institute, was organized,  and we started to go to Gatchina, but only on Thursday having the seminar on Mondays in the city. So, you see, that the communications were  hampered. Fortunately, for all physicists of the institute, the experimentalists  found the solution, organizing a regular Winter school. The idea was simple and brilliant: we go for a week in February-March outside of the town for skiing and physics. The theoretician would teach the experimentalists, what has happened in high energy physics during the past year,  and they would be able to ask whatever they want.  The scheme worked perfectly and 53-th   school will be held on 2-8 March 2019.
   
   The first school was in 1966, but we started to come to the school a bit later. It happened very fast, that the school became popular in the Soviet Union and many physicists from other institutes started to come to the school. Naturally, it was  quickly  transformed in the kind of informal conference or, better to say, workshop: the place where we, theoretical physicists, could discuss our problems with the guests from all around of the Soviet Union. However, the
   task to teach remained and large number of talks, devoted to the different aspects of nuclear and high energy physics, were read and gave us the possibility to be educated outside of the problems, that we worked with.
   
   Lev Lipatov transformed this school to the place where he presented his ideas and advocated his way of doing physics. He did this in  a very unusual and genius way: he organized his own seminar, which we called Lipatnik. It was outside of official program and   at a  specific time: after  supper and the night cultural program. The time was unlimited and very often he finished the seminar at 3 a.m. The main difference of this seminar from  others was  that Lipatov did not discuss the ideas and the way how to realized these ideas: he discussed just how to work out the ideas putting the main stress on how to calculate. 
  
  The participants felt more like students in the lecture but he was so skillful in giving his presentation, that everyone took part in the process. The benefits of such approach were obvious and the seminar was very popular. Actually, Lev had only one Ph.D. student: Ian Balitsky, but with these seminars he taught a lot of  students and young researchers. I believe, this seminar was the foundation of, so called, Lipatov school.
  
    \begin{figure}[h]
   \begin{center}
   \begin{tabular}{c c}
 \leavevmode
    \includegraphics[width=6.2cm]{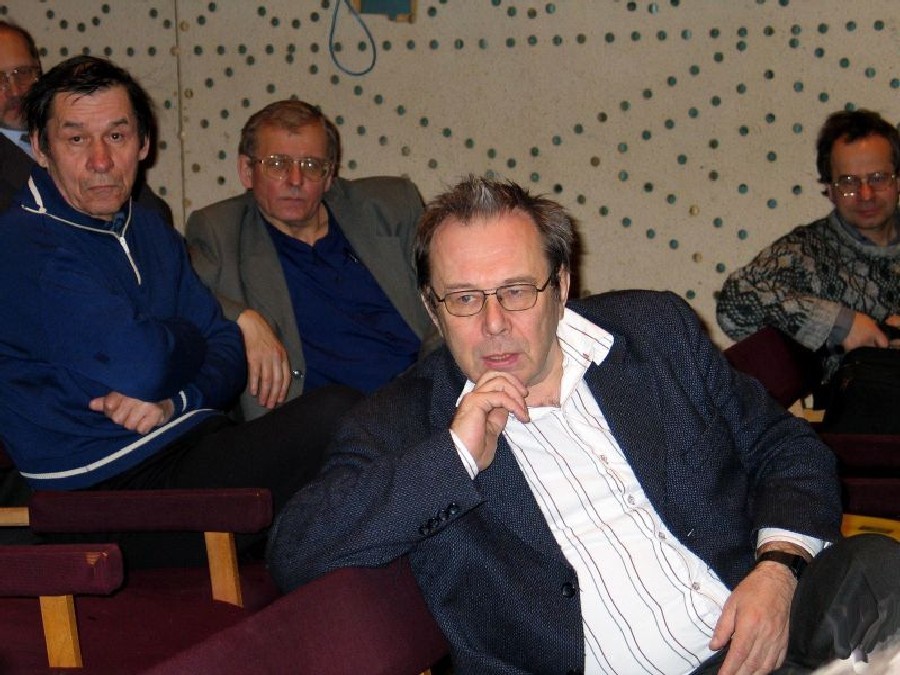} &  \includegraphics[width=6.cm]{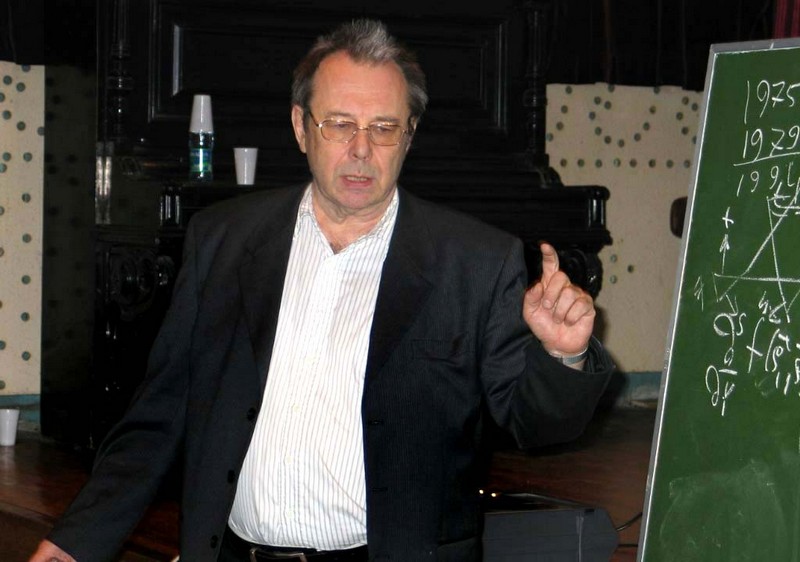}\\
    \ref{levsch}-a&   \ref{levsch}-b\\
    \end{tabular}
            \end{center}
    \caption{Lev Lipatov at the Winter  school in 2005 (\ref{levsch}-a) and in 2006 (\ref{levsch}-b).  You can see how he can listen and can discuss.}
    \label{levsch}
   \end{figure}
  As far as Winter schools are concerned, Lev felt that his duty to attend and to take part in preparation and in conducting the school. I put in this note  four photos of him. In the first two you see him in 2005 and 2006. Just to give you a flavour both of his participation and the problems that were discussed in the school I put below the titles of his talks at 39-th school:

1.  Kuraev E.A., Cherednikov I (JINR, Dubna), Antonov E.N, Lipatov L.N (PNPI) - Feynman diagrams for the effective
field theory at high energies in QCD;

2. Lipatov L.N.(PNPI) - Solution of Balitsky-Kovchegov 
equation in three-dimensional Yang-Mills theory;

3. Lipatov L.N., Volkov G., Velizhanin V.N.(PNPI),
Sabio Vera A.(ITP, Hamburg, Germany) -  
Calabi-Yau spaces,applications and classification.

   I hope you see the diversity of the problem and the serious attitude, that Lev had at these schools.

\begin{figure}[h]
   \begin{center}
   \begin{tabular}{c c}
 \leavevmode
    \includegraphics[width=6.35cm]{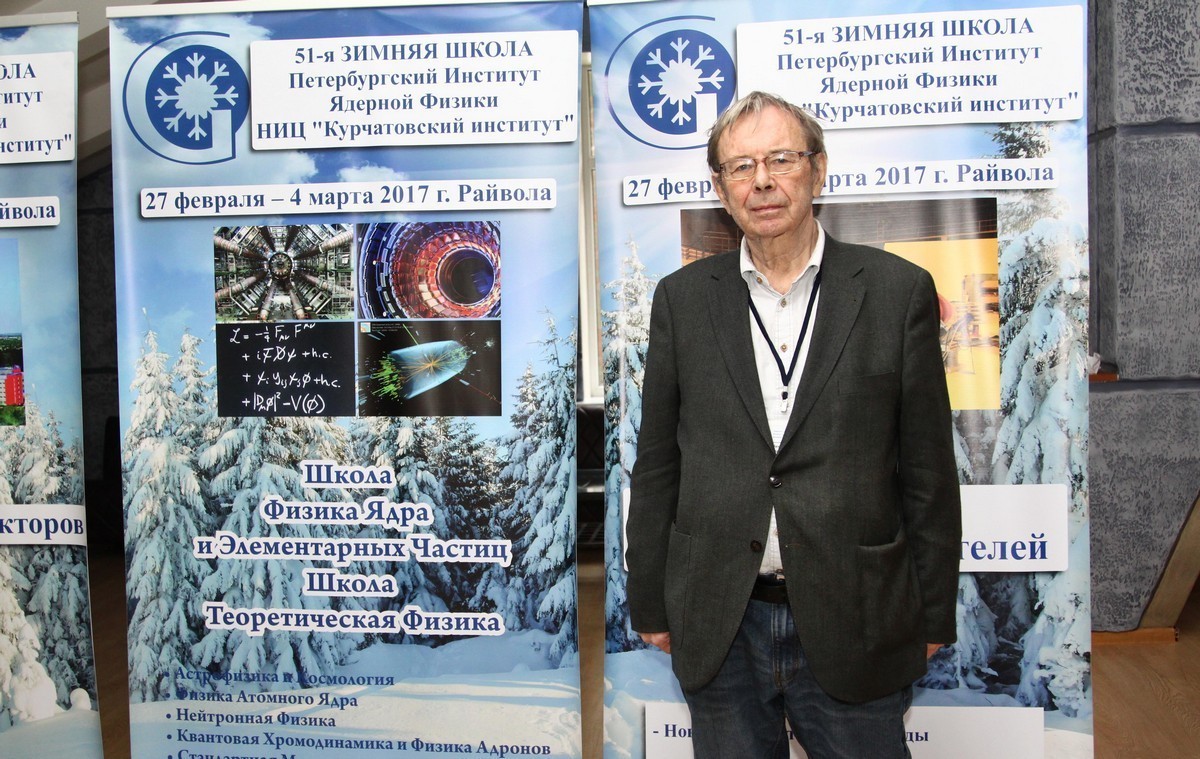} &  \includegraphics[width=6cm]{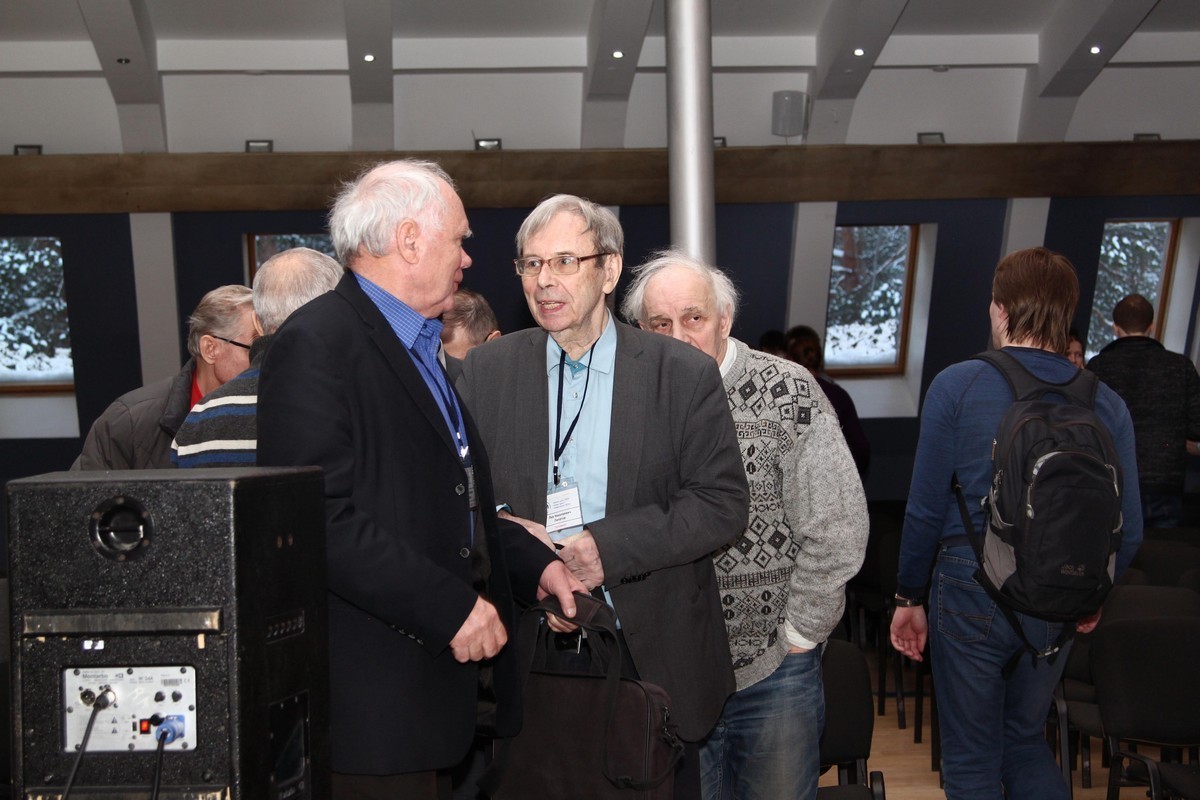}\\
    \ref{levsch1}-a&   \ref{levsch1}-b\\
    \end{tabular}
            \end{center}
    \caption{Lev Lipatov at his last school in 2017.}
    \label{levsch1}
   \end{figure}

     In Fig.\ref{levsch1}  you can see Lev during his last school in 2017. Fig.\ref{levsch1}-a shows him in front of the chart, where is written in Russian: The 51-th Winter School of Petersburg Nuclear Physics Institute,  February 27 - March 4,2017.    In Fig.\ref{levsch1}-b
    Lev and Victor Fadin are discussing something during this school. 

The title of his last talk was :``Scattering amplitudes in QCD and gravity". He made the Winter school a place, where every young theoretical physicists wished to participate. I have sent as many as possible  my students to him. All of them came back saying that this school were very useful for them. Therefore, Prof. Lev Lipatov made a considerable contribution to Israeli physics as well as to the Russian one.

  \section{BFKL Pomeron and me}
 
As I have alluded, I was doing mainstream  physics before the BFKL Pomeron time: the additive quark model (with L. Frankfurt), the reggeon calculus (with V. Gribov and A. Migdal) and duality models and hadronic strings (with V. Kudryavtsev). Starting from 1967 I had  possibility to work with Misha Ryskin on the processes of multiparticle generation: multiperipheral and parton models for the Reggeon (mostly Pomeron) structure. Misha is a physicist of the Lipatov's scale and to work with him was a god's gift to me. By 1974 we were disappointed by the theoretical background of the parton model and  were searching for the theory, that would solve the problems, that we had accumulated. Prof. Gribov advised us to look to the new QCD theory for this, saying that this theory looks as a good candidate for  nature.  We learned very fast the basics of the DLAP evolution equations but, fortunately, we also quickly realized, that the BFKL approach is our ticket to the new QCD physics. We found, that the structure of the parton cascade  of the BFKL approach is very similar to our familiar parton model  and we learned how to treat the differences.

Our first QCD  papers\cite{RYGL,LERYDY} were actually a breakthrough: we found that for the inclusive production in the BFKL cascade we have the $k_T$ factorization formula. We based these papers on Eq.\ref{GPROD}  (
the first diagram of Fig.\ref{bfkl})  and wrote the formulae for the  inclusive  process $h + h \to G(\gamma^*) + X$ in the form
\begin{equation}  \label{incl}
\frac{ d^2 \sigma}{ d y d^2 p_T}\,\propto \int d^2 k_T \frac{d  \sigma \left( GG \to G(\gamma^*)\right)}{d^2 p_T} \,\phi^G_h\left( Y - y; \vec{p}_T - \vec{k}_T\right)\,\phi^G_h\left(  y;  \vec{k}_T\right)
\end{equation}
where $\gamma^*$ stands for the virtual gamma quantum and $G$ for gluons.
$\phi$ is the probability to find  a gluon in a hadron which is the solution to the BFKL equation. 

Being newcomers to the field we did not realized neither the importance of this formula, nor the fact that we were first, who wrote it.  The papers, in spite of the fact that they  were translated to English, did not make any impact and the high energy community mostly was not aware about our results. Lev, who was the expert, never was interested in the multiparticle generation and he could not help us. We understood all these issues only ten years later and published the paper \cite{LRSS} in which has  everything, including the correct terminology. Being so slow, nevertheless, we were among the first people,  who suggested the $k_T$-factorization, but not the very first.  {\it Sic transit gloria mundi}.

Lev, when I told him, that his equation is suited for the estimates of the inclusive production at large values of transverse momenta, was very pleased and supportive. Such attitude encouraged me to tell him that actually his equation gives not the structure of the Pomeron, but the generalization of the DGLAP equation for the evolution of DIS at small values of  Bjorken $x$.  For the first ten minutes he told me how stupid I was, but for another ten minutes he was saying, maybe we are right. Our chat we finished with Lev telling, that he owes me a bottle of vodka, since he, finally, realized, that  we blew a new life in his beloved BFKL equation. We shared this bottle discussing how  beautiful  the BFKL equation  is.

Lev was also very supportive, when we  made the first attempt to include the BFKL evolution of the gluon density in  the GLAP equation (see our review \cite{GLR}) and take into account the running QCD coupling. However, he was lukewarm with our attempt to include the interaction of partons in the BFKL parton cascade by introducing the interaction of several parton showers. These ideas were taken from our previous experience with the Pomeron calculus  and the proof, which we gave, that the interaction of the BFKL Pomerons  can be treated in a similar fashion as  is done in the Gribov Pomeron calculus.

He certainly knew the two major problems with the BFKL Pomeron: power-like increase with energy, and the influence of the confinement of quarks and gluon on the energy evolution.
 We showed, that the interaction of the BFKL Pomeron leads to the saturation of the gluon density, which stops  increasing, and to the transverse momentum in the energy evolution, which does not decrease. The typical transverse momentum increases, making the approach better and better from the point of the perturbative estimates. In spite of  his numerous attempts to find a hole in our proof,
he failed. However, his intuition did not trust our approach to the end of his days. As I have mentioned, he hoped to solve these problems using the effective action, which he suggested. In simple words, he wanted to include the parton annihilation directly to the BFKL kernel. 

Actually, Misha and I  knew, that  the non-linear evolution equation that we suggested , did not lead to the parton cascade, that satisfies the $s$-channel unitarity at any values of  rapidity of gluons. It is written clearly in our paper\cite{GLR}. However, we 
believed, that the equation is only valid in the vicinity of the new dimensional scale: saturation momentum, where it  correctly  introduces the gluon interaction. Lev thought that we need to solve exactly and for entire region of the rapidity.

  I have a history of my understanding of his ideas. In the beginning, as I have mentioned,  I believed that our approach and the non-linear equation, which we suggested is only valid in the vicinity of the  saturation momentum. Inside the saturation domain we cannot trust the perturbative approach to QCD, if not QCD itself. We suggested just to assume,  that the density reaches some value as the practical way to introduce unsolved confinement into our evolution.

  When Balitsky\cite{BAL} and Kovchegov\cite{KOV}, followed the ideas of Al Mueller\cite{MUCD}, re-wrote our equation in the coordinate representation and clarified the physical meaning of $\phi$, relating it to the dipole scattering amplitude, they managed to convinced me, that the resulting equation is valid in the wide kinematic region. This claim was in the direct contradiction with Lev's expectation, but since it was done in the dipole approach, Lev refused even discuss the issue with me. 
  
  However, later Al Mueller and A. Shoshi\cite{MUSH} showed that the non-linear equation even in  the form,  that Balitsky and Kovchegov derived, does not provide  $s$-channel unitarity for colourless dipoles with any value of its rapidity, reviving our old arguments. Starting from 2004,  I went 
   back in my disbelief, that this equation is correct for any values of energy.   
   In recent paper\cite{KLL} we found, that this equation contradicts the $s$-channel unitarity and  it is deadly sick.  At the moment, I do not know how the contradictions, that we found in Ref.\cite{KLL} are interrelated with Mueller and Shoshi arguments, but I hope to find the generalization of this equation
with help of Alex Kovner and Misha Lublinsky and, perhaps, that Lev's ideas will help us in this adventure. The simple model, which has the same form and the same defects as the Balitsky-Kovchegov equation, and, which we solved in the paper, showed that the solution will be  non-trivial. It is interesting to note, that this paper has no citations and in a sense situation looks as with our papers of 1980. I consider, this as an indication, that something interesting is awaiting  us in the future. 

\begin{figure}[h]
   \begin{center}
 \leavevmode
    \includegraphics[width=8cm]{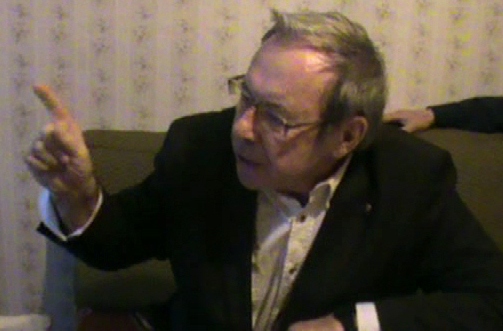}\\ 
    \end{center}
    \caption{Lev Lipatov discussing with me his effective action in 2013.}
    \label{levD}
   \end{figure}

Looking at the picture in Fig.\ref{levD}, I imagine that Lev is instructing me, how to solve the problem. Actually, I know what he would tell: as I told you many times, you need to use my effective action. Perhaps, he is right, but I will try something less complicated.

Having so different points of view on the non-linear evolution, we could not find a subject for joint work for  a long time. However, rather recently, we did one\cite{LILE}. The problem, which we address in this paper,  rooted in 1980.  During our first talk at the Gribov seminar,  we were attacked  with the question, what happens with our non-linear corrections at large impact parameters , and how they can heal the problem that the BFKL Pomeron has the power - like decrease at large $b$. Prof. V. Anisovich was a leader of these attacks. Since we knew this problem, we were able only to tell them, that we propose our equation for the gluon densities at $t=0$ and proved that we can take the integration over  the momentum transferred by the BFKL Pomerons,  introducing only one new phenomenological parameter: the correlation length of gluons in the parton cascade. Certainly, we were aware that this is not satisfactory answer.  We made several attempts to introduces a new non-perturbative soft parameter to describe the behaviour at large $b$ (see Refs.\cite{LERY0,LERY1,LETA,LEPION} , but we failed to find  a non-ambiguous, theoretical way to introduce a new non-perturbative parameter with clear physical meaning, in the kernel of the BFKL equation. The non-linear Balitsky-Kovchegov equation in the coordinate representation is written at fixed impact parameter. Both Ian and Yura considered this as an essential difference from our approach. When I heard this for the first time, I just laughed: kids did not understand what pandora's box they opened. Actually, I agree with them, that two equations are different, but ours is better, since it shows a way to avoid the problem of large $b$ behaviour at least for the deep inelastic structure functions.
 Indeed, very soon after, A. Kovner and U. Wiedemann\cite{KW} 
made  public this secret and the large $b$ dependence is the most important unsolved problem which prevents the CGC approach from  being  an effective theory for QCD at high energies.

So, it is quite natural that both Lev and I  tried to joint  our efforts to think about this.
In our paper with Lev we suggested to consider not QCD, but the gauge theory with the same colour structure as QCD , but with Higgs mechanism for the mass of gluons. This theory has no problem with the behaviour of the scattering amplitude at large $b$,  but, at first sight, we changed the eigen functions and eigenvalues of the Pomeron in this theory in comparison with QCD. However, 
 we found that the intercept of the Pomeron is the same as the intercept of the BFKL Pomeron in QCD. We consider this fact as an indication, that introducing correction from confinement of quarks and gluons in the kernel of the BFKL equation will not change the main prediction of the CGC approach. On the other hand, Victor Fadin  poured  a bucket of cold water  on us claiming that this result is trivial. I am still waiting for his paper in which he will prove, that all possible modifications of the BFKL kernel will give the same spectrum.

\section{The best papers}

I believe, that without  an exaggeration,  we can say that all our understanding of high energy scattering in QCD is based on Lev Lipatov and his team's  papers. Even more, perhaps, they will live longer than  his other  contributions, but he himself as well as many other physicists, consider that his best papers are  the ones on high order of the perturbative series\cite{LIPHO1,LIPHO2,LIPHO3,LIPHO4,LIPHO5}. Indeed, the method to calculate the asymptotic behaviour even in the BFKL kinematic, was suggested in the paper with Frolov and Gribov, and Lev always remembered , that he was a developer but not a creator. The problem of the  behaviour of the large order of the perturbative estimates and the divergence of the perturbative series was started by him, was formulated and solved. Not being an expert in this field I took the following lines from the obituary of  the theory division of PNPI  faculty and staff:

``  Lev Nikolaevich Lipatov excelled in research of the perturbative Quantum Field Theory. He had an amazing ability to find elegant and mathematically rigorous solutions of prohibitively difficult problems. He had suggested and developed an effective method for analysis of the asymptotic behavior of the perturbation theory series in Quantum Field Theory. His works on this subject  now acquired  the status of the theoretical physics classics. They now form the basis of our knowledge about the behavior of the perturbation theory series and their singularities in the complex coupling constant plane. "

In these papers Lev showed that he is more than the expert in high energy QCD. He became an internationally  renowned theoretical physicist with deep knowledge  all methods of the modern theoretical physics. He became a  friend of the   Chernogolovka theory group of Sasha Polyakov and, when most of this group moved to Paris and other places in the West, he frequently visited them, discussing problems, that are far away from  high energy scattering.

\section{1990-2017: climbing the ladder of success}

\begin{figure}[h]
   \begin{center}
 \leavevmode
    \includegraphics[width=8cm]{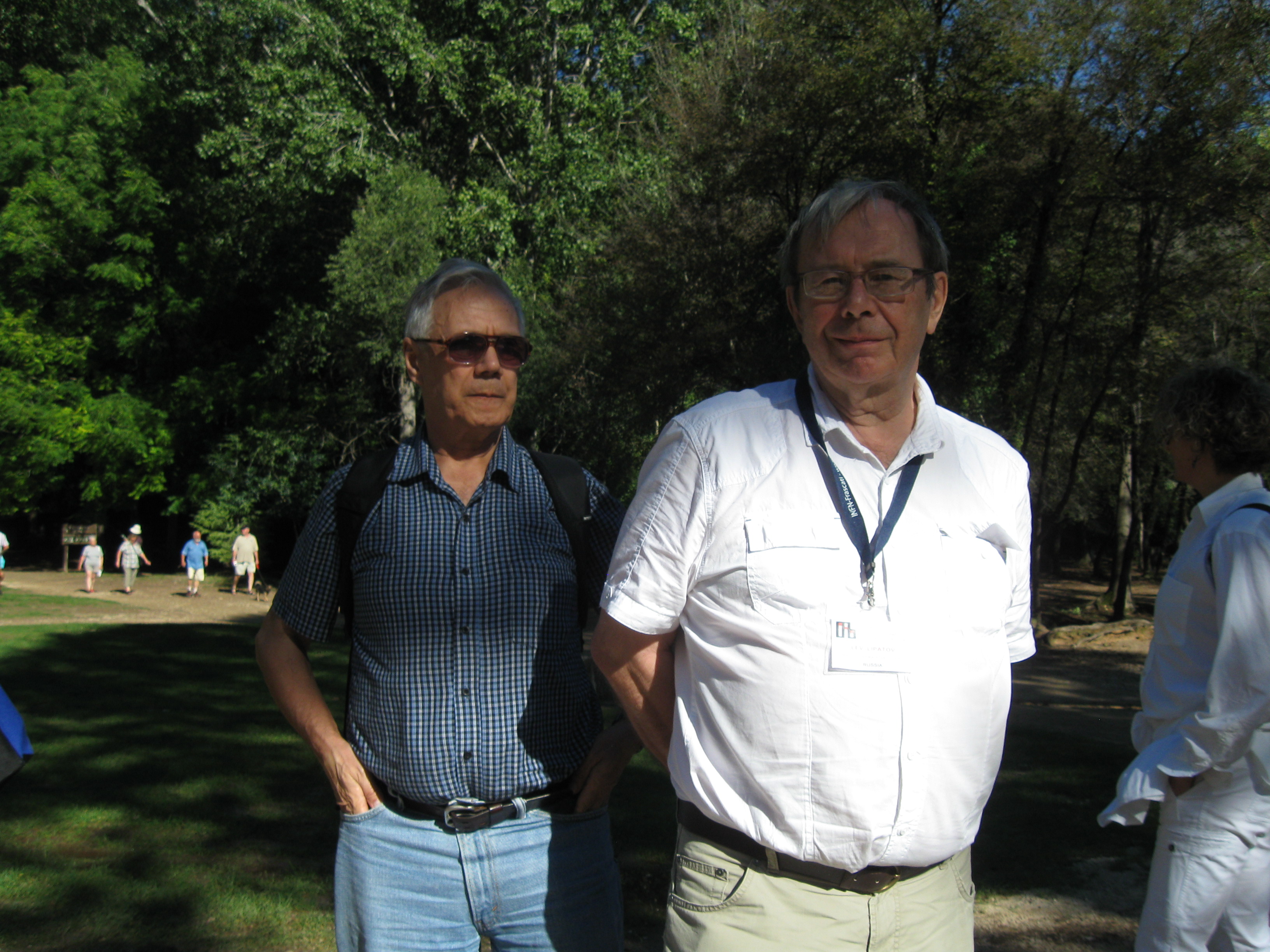}
    \end{center}
    \caption{Lev Lipatov  and me at Diffraction 2014., Primosten, Croatia.}
    \label{lippr}
   \end{figure}

\begin{figure}[h]
   \begin{center}
 \leavevmode
    \includegraphics[width=8cm]{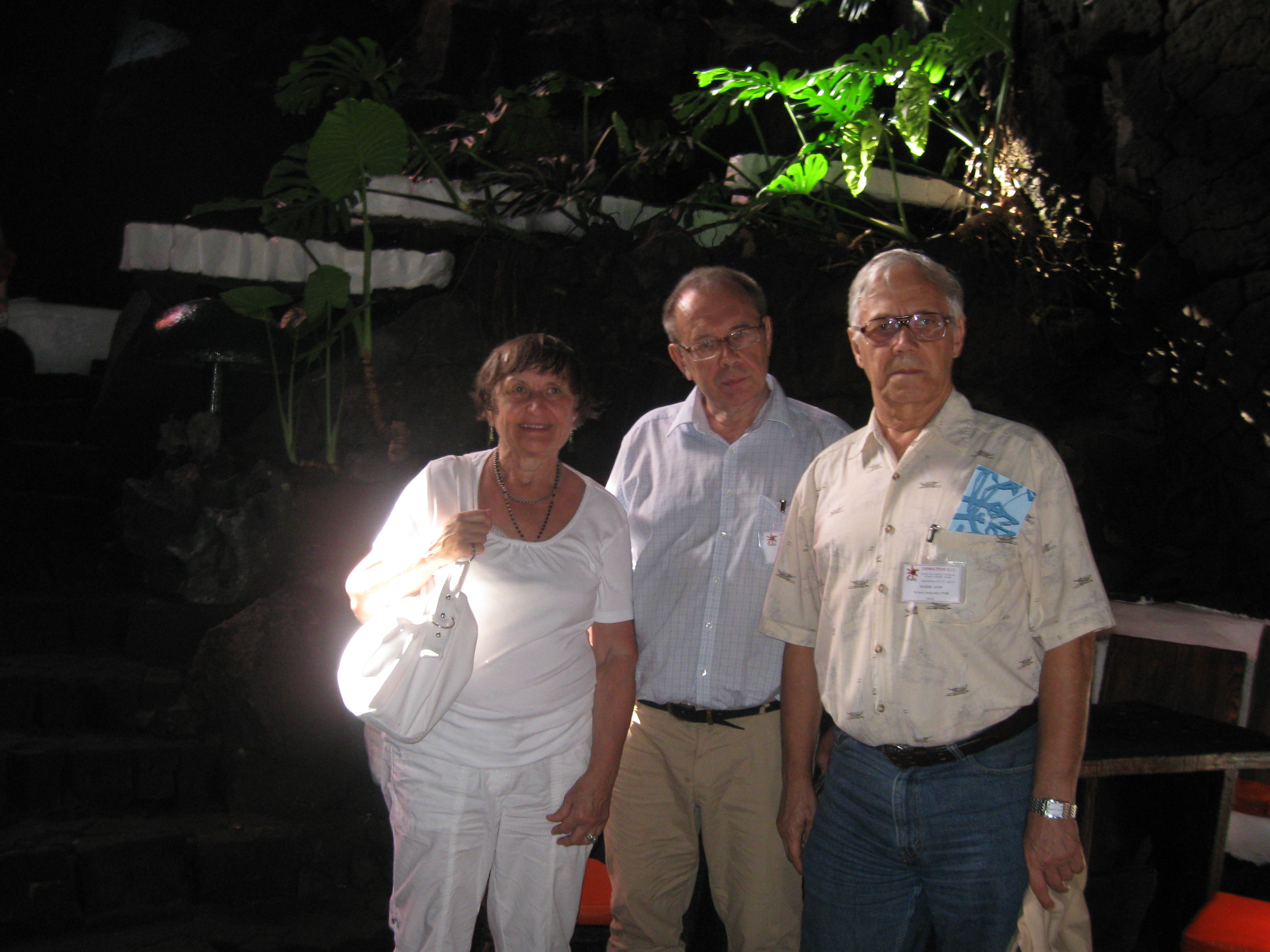}
    \end{center}
    \caption{Lev Lipatov , his wife, Elvira, and me at Diffraction 2012., Lanzarote, Canary Islands, Spain.}
    \label{liplan}
   \end{figure}

\begin{figure}[h]
   \begin{center}
 \leavevmode
    \includegraphics[width=8cm]{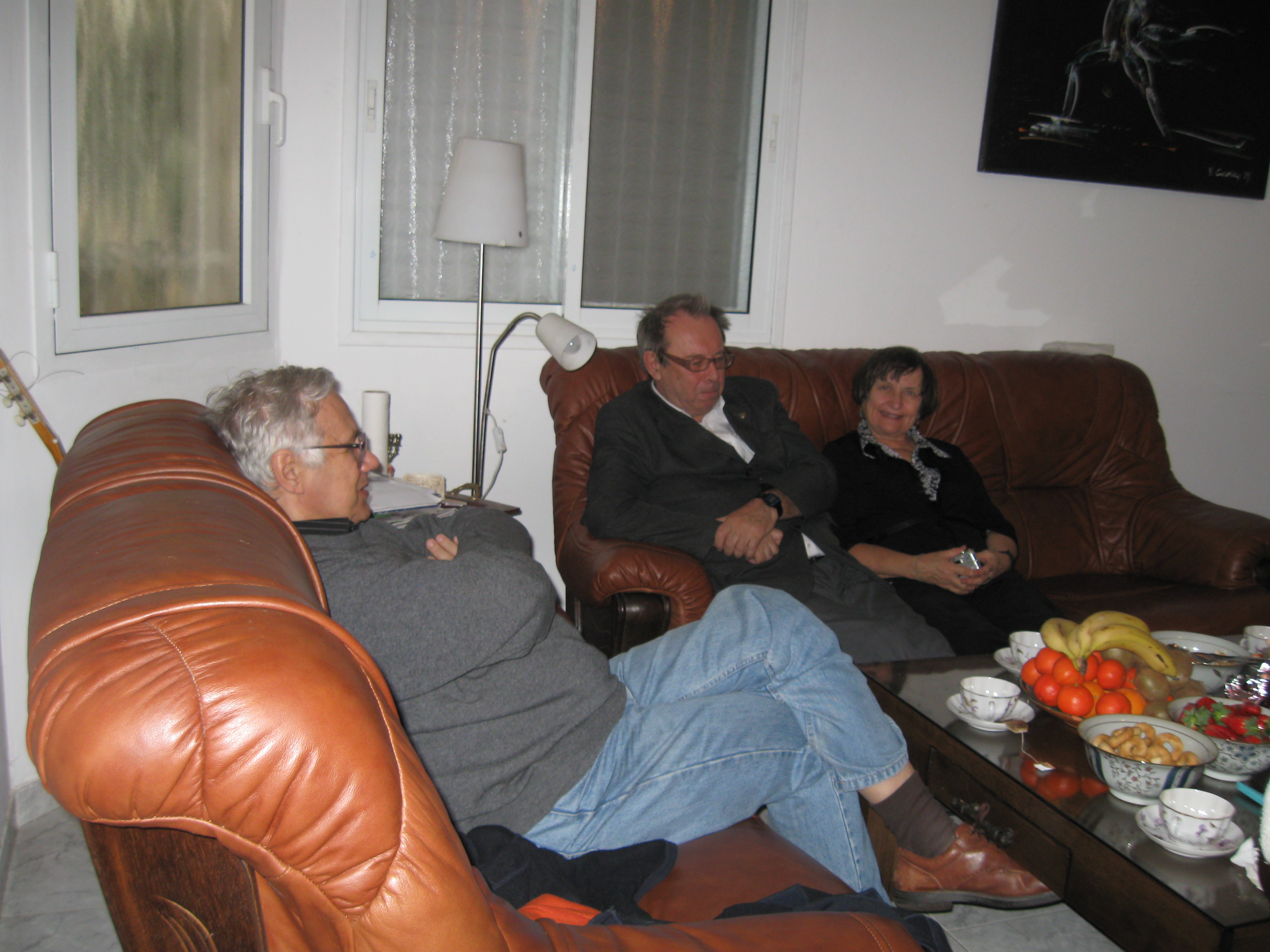}\\ 
    \end{center}
    \caption{Lev Lipatov and his wife, Elvira,  with me   in my home at Be'er Yakov, Israel.}
    \label{lipmh}
   \end{figure}

In 1990 I left Russia and in 1996 I realized the dream of my teacher, Prof. V. Gribov, and became a member of particle physics department of Tel Aviv University. I remember, as it was yesterday, as once in 1964 or 1965  Prof. Gribov  came to the common room saying:``Guys, now the center of the theoretical physics moves from Caltech and Harvard to Tel Aviv. Yuval Ne'eman organizes a  high energy group  at the new university there".  He did not say anything else,  but sighed heavily. Of course, he exaggerated a bit, but every time, when my student comes to my office, opening the door with his foot, and tells me, without saying ``Hi": ``Professor, you are stupid, everything that you told me yesterday, is just rubbish", I feel that Tel Aviv is a good place for doing physics with a  bright future ahead. Hence, I could watch only  from the side, how my friend went from  one success to another, but  onlooker sees most of the game. Especially, because Lev started to travel abroad very frequently and we met both at the conferences and at the different institutions (see  Fig\ref{lippr} and Fig\ref{liplan}). It goes without saying that I used every possibility to invite him to Tel Aviv and to the  conferences in Israel, where we can talk in very informal atmosphere of my home (see Fig.\ref{lipmh}).

  During teaching, I have tried to use Lev for training my students. I  remember him saying: `` I cannot teach students how to think, but I can teach  even a child how to calculate". I tried  for every  student of mine  to organize either the possibility to work with him , or to meet him and discuss physics during as long as possible period of time.  I used to send my students to DESY, to Prof. J. Bartels since I know that Lev visits him on more or less permanent basis. For three of my students the cooperation with Lev resulted in joint papers: nine for  Alex Prygarin, three for Andrey Kormilitzin and two for Sergey Bondarenko. After my retirement, Bondarenko and Prygarin found money  for Lev's visits giving me a chance to see my friend (see Fig.\ref{lipmh}). By the way, the last Lev's papers were written with them.

  During almost three decades Lev made several breakthrough, but one of them: the calculation the BFKL equation in the next-to-leading order\cite{FALINLO}, impressed me enormously. I believe, that Victor Fadin and he accomplished the feat  of performing the needed calculations. It took them almost ten years to find the answer, and I am sure,  that they did this only because Jochen Bartels helped them organizing  the regular visits to Hamburg for them  and convincing them to continue these, time consuming, calculations. I think, that the high energy community should be very thankful to Jochen. The  NLO kernel for the BFKL equation led to deeper understanding the interrelation between BFKL and DGLAP evolutions \cite{SAL,CCS}. On the other hand, the fact that the NLO corrections turn out to be large resolve the contradiction with the attempt to describe the experimental data in the framework of the BFKL evolution or/and in the framework of the CGC/saturation approach.  Indeed, the two essential parameters that determine the high energy scattering are the BFKL Pomeron intercept, which is equal to $ \approx\,\,2.8 \alpha_S$ and the energy behaviour of the new dimensional scale: saturation momentum  $Q^2_s \propto \exp\left(4.88\,\alpha_S \ln(\frac{1}{x})\right)$ . Both of them show the increase in the leading order BFKL approach, which cannot be reconciled with the available experimental data. So, the large NLO corrections looks as the only way out,  even now as well as  two decades ago.

  I tried to follow  other achievements of my friend especially his proof that the BFKL Hamiltonian is equivalent to an integrable Heisenberg spin
model which introduced the concept of integrability to the high energy physics. He developed further his effective action approach and actually, his lecture on this subject was the last words, that I heard from him\cite{LIPEA}. For a couple of years, I was very enthusiastic about AdS/CFT duality and
  discussed a lot with Lev his   duality between BFKL Pomeron and the graviton. However, I felt disappointed and cheated, since this picture in which the BFKL Pomeron mostly contributes to the real part of the scattering amplitude, contradicts everything that we have learned both experimentally and theoretically about the processes of multi-particle generation.  Fortunately, physics is an experimental science and, I firmly believe, that the LHC data on soft interactions killed these ideas since they do not show any significant increase of  diffraction dissociation cross sections with the growth of energy.

 It goes without saying, that his scientific achievements got  wide recognition resulting in his advancing in rank. In 1990 he became ta Professor of  Physics and from 1996 to the end of his days  he was the head of the Gribov department at Petersburg Nuclear Physics Institute.
 In 1997 he was elected   a  corresponding member of the Russian Academy of Science. Lev Lipatov was a great patriot of Russia. His  dream from the student years was to become the Russian academician. This dream  was accomplished in 2011, when he became a full member of 
 the Russian Academy of Science.
 
  I wish to make a 
 lyrical digression in my narration about his life. I just want to tell my friends from the West that Russia is very complicated country, but, please, think about Prof. Lipatov,  but not about politicians, when you discuss Russians. Actually, he was against my move to Tel Aviv and discussed with me a number of arguments, why I need to stay in Russia. Actually, all these arguments were correct, but I failed to convince him, that it is difficult to live knowing, that you need to be the best to survive. It is funny, but he had only one argument for, he told me, that he envy me, because I will have extraordinary bright students  in Tel Aviv. Indeed, in Russia, the Jewish students were on the top of the class. To my surprise, he was wrong in his expectation, the students in Tel Aviv turned out to be  the same as Russian students, even they were lazy like Russians. So antisemitism has a positive impact.
 
 Since, the differences in our opinions were not of the scale of the gluon reggeization, we melted these differences and Lev was a frequent guest in Tel Aviv. In  2009-2010 he was  scholar at Sackler Institute of Advance Studies and spent almost one year with us. He told me, that he likes Israel, since it is not a boring country. During the last three or four years he has come to Israel to Ariel university to work with my former students allowing  me the  pleasure to talk with him.
   
   I think, it is enough to list  the prizes and awards that he got, to illustrate what we lost:
   
   1. 1993- the Research Award of the Alexander- von-Humboldt
Foundation. He spent  more than one year in DESY.
   
\begin{figure}[h]
   \begin{center}
 \leavevmode
    \includegraphics[width=7cm]{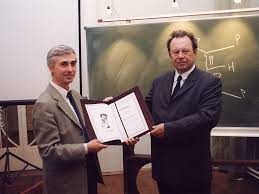}\\ 
    \end{center}
    \caption{2001: Lev Lipatov is getting the Pomeranchuk prize  from Mikhail Danilov, the director of  ITEP .}
    \label{levpompr}
   \end{figure}

   2. 2001-    I. Ya. Pomeranchuk Prize (2001) for the BFKL Pomeron (see Fig.\ref{levpompr});
   
 3. 2006-2009: the Marie Curie ExcellenceChair of the European Community, hosted by Hamburg University;
 
 4. 2009-2010:  Scholar at Sackler Institute of Advance Studies (Tel Aviv);,
 
  5. 2015:   European Physical Society High Energy and Particle Physics Prize 
 `` for developing a probabilistic field theory framework for the dynamics of quarks and gluons, enabling a quantitative understanding of high-energy collisions involving hadrons" , or in simple words, for 
   the DGLAP evolution equation;

   \section{Life in numbers}
    
    The content of this section is just the table below which says for itself without any comment.

    \begin{table}[h]
\begin{tabular}{|l  l|}
\hline
 Number of papers & 160\\  
 Number of citations & 26 169\\
 Citations per paper: &170.9\\
 Papers with citation 3000-4000 & 3\,\, (Refs. \cite{DGLAP,FKL3,BALI})\\
  Papers with citation 1000-3000 & 6\\
   Papers with citation 500-1000 & 1\\
   Famous papers (250-499) &  8\\
Very well-known papers (100-249) &18\\
Well-known papers (50-99) &  21 \\
Reviews &6\\
Books & 3\\
Number of co-authors for $\geq 3$ papers & 36\\
Total number of co-authors: & 86\\
\hline
 \end{tabular}
\label{t1}
\end{table}
   I believe that these numbers tell us more about the main contributions that Lev Lipatov did to  high energy physics, than the prizes and awards.
    \begin{figure}[h]
   \begin{center}
 \leavevmode
    \includegraphics[width=7cm]{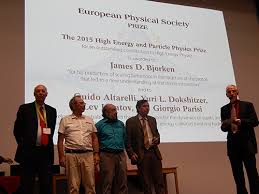}\\ 
    \end{center}
    \caption{2015:  DGLAP(without Gribov): Dokshitzer, Lipatov, Altarelli and Parisi, are  getting the european physical society prize.}
    \label{leveurpr}
   \end{figure}
   \section{My several words}
   Every day, when I start to work, I feel myself as in the Plato's cave: your faces, my dear friends: Sasha Vassiliev,  Alyosha Kaidalov, Lev Lipatov, Odette Benary, Uri Maor, Yona Oren, are on the wall in front of me. Thank you for sharing your life with me, thank you for your support and generosity, thank you for the intellectual pleasure of our discussions, which adorned my life.
  I wish to report you, that I am still doing physics, but, I must admit , that only nine hours per day.  
   I still enjoy doing physics, I think, that we made a correct  decision,  choosing physics as the way of life. At least I have not felt bored even a minute in my life.
   
    Every time when I put my formulae on the sheet of paper I feel myself as a young guy from Gribov's department, who is hurrying up to meet you and discuss, what I have understood.
   All our hopes, all our ideas, all our dreams are still with me and I am trying to share them with young physicists.

\begin{figure}[h]
   \begin{center}
 \leavevmode
    \includegraphics[width=7cm]{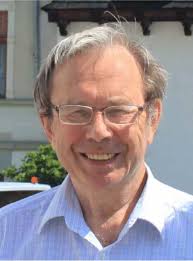}\\ 
    \end{center}
    \caption{ ``Vitam regit fortuna, non sapientia" (``Fortune, not wisdom, rules lives") ({\it Cicero}).}
    \label{last}
   \end{figure}

   Dear Lev, the renowned physicist and my friend for 60  years, my memory will not fade, I will remember you, our discussions, our approaches to the life and physics. You will be always as in Fig.\ref{last}: smiling, active, always in motion and always ahead of all us.

\begin{thebibliography}{99} \frenchspacing
\bibitem[{Dokshitzer(2004)}]{DOKH}
Y.~L.~Dokshitzer and D.~E.~Kharzeev,
  {\it ``The Gribov conception of quantum chromodynamics,''}
  Ann.\ Rev.\ Nucl.\ Part.\ Sci.\  {\bf 54} (2004) 487,
  [hep-ph/0404216].
  \bibitem[{Dokshitzer(1998)}]{DOK}
   Y.~L.~Dokshitzer,
  {\it ``V.N. Gribov: 1930 - 1997,''}
  Submitted to: Phys.World
  [physics/9801025].
  
  \bibitem[{Frolov(1966a)}]{DLA1}
  G.~V.~Frolov, V.~G.~Gorshkov and V.~N.~Gribov,
  {\it ``Double logarithmic asymptotics of the high-energy Compton scattering,''}
  Phys.\ Lett.\  {\bf 20} (1966) 544.
  \bibitem[{Frolov(1966b)}]{DLA2}
  G.~V.~Frolov, V.~G.~Gorshkov and V.~N.~Gribov,
 {\it  ``Backward Compton scattering at high-energies,''}
  Phys.\ Lett.\  {\bf 22} (1966) 662.
  \bibitem[{Gribov(1967)}]{DLA21}
   V.~N.~Gribov, V.~G.~Gorshkov and G.~V.~Frolov,
 {\it  ``Double logarithmic asymptotics of the large angle Compton scattering,''}
  Annals Phys.\  {\bf 43} (1967) 201.
    \bibitem[{Gorshkov(1968a)}]{DLA3}
  V.~G.~Gorshkov, V.~N.~Gribov, L.~N.~Lipatov and G.~V.~Frolov,
 {\it  ``Double logarithmic asymptotic behavior in quantum electrodynamics,''}
  Sov.\ J.\ Nucl.\ Phys.\  {\bf 6} (1968) 95
   [Yad.\ Fiz.\  {\bf 6} (1967) 129];  Phys.\ Lett.\  {\bf 22} (1966) 671.  
  \bibitem[{Gorshkov(1968b)}]{DLA4}
  V.~G.~Gorshkov, V.~N.~Gribov, L.~N.~Lipatov and G.~V.~Frolov,
 {\it  ``Backward electron - positron scattering at high-energies,''}
  Sov.\ J.\ Nucl.\ Phys.\  {\bf 6} (1968) 262
   [Yad.\ Fiz.\  {\bf 6} (1967) 361].
  \bibitem[{Gorshkov(1969)}]{DLA5}
   V.~G.~Gorshkov, L.~N.~Lipatov and M.~M.~Nesterov,
 {\it  ``On j-plane singularities of e-minus muon-minus scattering amplitudes,''}
  Yad.\ Fiz.\  {\bf 9} (1969) 1221.
  \bibitem[{Lipatov(1969)}]{DLA6}
  L.~N.~Lipatov and G.~V.~Frolov,
 {\it  ``Some processes in quantum electrodynamics at high energies and small scattering angles,''}
  Pisma Zh.\ Eksp.\ Teor.\ Fiz.\  {\bf 10} (1969) 399.
  \bibitem[{Anselm(1969a)}]{DLA7}
   A.~A.~Anselm, L.~N.~Lipatov and G.~A.~Winbow,
 {\it  ``Three-particle production amplitude at high ernergies and momentum transfers,''}
  Pisma Zh.\ Eksp.\ Teor.\ Fiz.\  {\bf 10} (1969) 499.
  \bibitem[{Anselm(1969b)}]{DLA8}
  A.~A.~Anselm, L.~N.~Lipatov and G.~A.~Winbow,
 {\it  ``Amplitude for three-particle production at high energies and large momentum transfers,''}
  Nuovo Cim.\ A {\bf 69} (1970) 57.
  \bibitem[{Lipatov(1971a)}]{DLA9}
   L.~N.~Lipatov,
 {\it  ``Bremsstrahlung in $e^+\, e^-$ backward scattering at high energies.,''}
  Yad.\ Fiz.\  {\bf 14}, 396 (1971).
  \bibitem[{Lipatov(1971b)}]{DLA10} 
  L.~N.~Lipatov,
  {\it ``Bremsstrahlung in backward $e^+\, e^-$- scattering as a multiregge process,''}
  Yad.\ Fiz.\  {\bf 14} (1971) 160.
  \bibitem[{Lipatov(1971c)}]{DLA11}
   L.~N.~Lipatov and G.~V.~Frolov,
 {\it  ``Some processes in quantum electrodynamics at high energies.,''}
  Yad.\ Fiz.\  {\bf 13} (1971) 588.
  
  \bibitem[{Gorshkov(1971)}]{DLA12}
   V.~G.~Gorshkov, E.~A.~Kuraev, L.~N.~Lipatov and M.~M.~Nesterov,
  {\it ``Interference between coulomb and strong interactions at high energies,''}
  Zh.\ Eksp.\ Teor.\ Fiz.\  {\bf 60} (1971) 1211.
  \bibitem[{Kuraev(1972a)}]{DLA13}
  E.~A.~Kuraev and L.~N.~Lipatov,
  {\it ``Total cross-section for electron and muon pair production in colliding electron beams,''}
  Pisma Zh.\ Eksp.\ Teor.\ Fiz.\  {\bf 15} (1972) 229.
 \bibitem[{Kuraev(1972b)}]{DLA14}
  E.~A.~Kuraev and L.~N.~Lipatov,
 {\it  ``Electron and muonic production in $e^+\, e^+$ and $e^+\, e^-$  colliding beams,''}
  Yad.\ Fiz.\  {\bf 16} (1972) 1060.
   \bibitem[{Kuraev(1973a)}]{DLA15}
  E.~A.~Kurayev, L.~N.~Lipatov, N.~P.~Merenkov and V.~S.~Fadin,
 {\it  ``Radiative corrections to bremsstrahlung radiation at high energies,''}
  Zh.\ Eksp.\ Teor.\ Fiz.\  {\bf 65} (1973) 2155.
  \bibitem[{Kuraev(1973b)}]{DLA16}
   E.~A.~Kuraev, L.~N.~Lipatov and N.~P.~Merenkov,
 {\it  ``High energy behavior of total cross-sections for $e^+\, e^-$- scattering and the slope at q-squared = 0 of the Dirac form-factor,''}
  Phys.\ Lett.\  {\bf 47B} (1973) 33.
  \bibitem[{Kuraev(1974a)}]{DLA17}
  E.~A.~Kuraev, L.~N.~Lipatov, N.~P.~Merenkov, V.~S.~Fadin and V.~A.~Khoze,
  {\it ``Double bremsstrahlung in the same direction in $e^+\, e^+$ colliding beams,''}
  Sov.\ J.\ Nucl.\ Phys.\  {\bf 19} (1974) 164
   [Yad.\ Fiz.\  {\bf 19} (1974) 331].
   \bibitem[{Kuraev(1974b)}]{DLA18}
  E.~A.~Kuraev and L.~N.~Lipatov,
 {\it  ``Bremsstrahlung mechanism of e+ e- and mu+ mu- pair production in electron colliding beams,''}
  Yad.\ Fiz.\  {\bf 20} (1974) 112.
  \bibitem[{Kuraev(1974c)}]{DLA19}
   E.~A.~Kuraev, L.~N.~Lipatov and M.~I.~Strikman,
 {\it  ``On the relation between various inclusive processes in quantum electrodynamics,''}
  Zh.\ Eksp.\ Teor.\ Fiz.\  {\bf 66} (1974) 838.
  \bibitem[{Drell(1969)}]{DRELL}
   S.~D.~Drell, D.~J.~Levy and T.~M.~Yan,
 {\it  ``A Theory of Deep Inelastic Lepton-Nucleon Scattering and Lepton Pair Annihilation Processes,''}
  Phys.\ Rev.\  {\bf 187}, 2159 (1969);   Phys.\ Rev.\ D {\bf 1} (1970) 1035,1617,2402.
  \bibitem[{Gribov(1971)}]{DGLAP}
   V.~N.~Gribov and L.~N.~Lipatov, 
{\it  ``Deep inelastic $ e p$ scattering in perturbation theory,''}
  Sov.\ J.\ Nucl.\ Phys.\  {\bf 15} (1972) 438
   [Yad.\ Fiz.\  {\bf 15} (1972) 781];\,\,\,
   
   \bibitem[{Gribov(1972)}]{DGLAP1}
   V.~N.~Gribov and L.~N.~Lipatov,   {\it  ``$e^+\, e^+$ pair annihilation and deep inelastic $ e p$ scattering in perturbation theory,''}
  Sov.\ J.\ Nucl.\ Phys.\  {\bf 15} (1972) 675
   [Yad.\ Fiz.\  {\bf 15} (1972) 1218].
  \bibitem[{Dokshitzer(1977)}]{DOKSH}
   Y.~L.~Dokshitzer,
  {\it ``Calculation of the Structure Functions for Deep Inelastic Scattering and e+ e- Annihilation by Perturbation Theory in Quantum Chromodynamics.,''}
  Sov.\ Phys.\ JETP {\bf 46} (1977) 641
   [Zh.\ Eksp.\ Teor.\ Fiz.\  {\bf 73} (1977) 1216].
  \bibitem[{Altarelli(1977)}]{AP}
   G.~Altarelli and G.~Parisi,
  {\it ``Asymptotic Freedom in Parton Language,''}
  Nucl.\ Phys.\ B {\bf 126} (1977) 298.
  \bibitem[{Christ(1972)}]{MUDIS}
   N.~H.~Christ, B.~Hasslacher and A.~H.~Mueller,
  {\it ``Light cone behavior of perturbation theory,''}
  Phys.\ Rev.\ D {\bf 6} (1972) 3543.
  \bibitem[{Lipatov(1974)}]{LIPDIS}
  L.~N.~Lipatov,
  {\it ``The parton model and perturbation theory,''}
  Sov.\ J.\ Nucl.\ Phys.\  {\bf 20} (1975) 94
   [Yad.\ Fiz.\  {\bf 20} (1974) 181].
  \bibitem[{Gribov(1971)}]{GLF}
    V.~N.~Gribov, L.~N.~Lipatov and G.~V.~Frolov,
   {\it``The leading singularity in the $ j $-plane in quantum electrodynamics,''}
  Sov.\ J.\ Nucl.\ Phys.\  {\bf 12} (1971) 543
   [Yad.\ Fiz.\  {\bf 12} (1970) 994];\,\,\,Phys.\ Lett.\  {\bf 31B} (1970) 34. 
   \bibitem[{Fadin(1975)}]{FKL1}
  V.~S.~Fadin, E.~A.~Kuraev and L.~N.~Lipatov,
 {\it ``On the Pomeranchuk Singularity in Asymptotically Free Theories,''}
  Phys.\ Lett.\  {\bf 60B} (1975) 50.
  \bibitem[{Lipatov(1976)}]{LIPBFKL}
    L.~N.~Lipatov,
{\it ``Reggeization of the Vector Meson and the Vacuum Singularity in Nonabelian Gauge Theories,''}
  Sov.\ J.\ Nucl.\ Phys.\  {\bf 23} (1976) 338
   [Yad.\ Fiz.\  {\bf 23} (1976) 642]. 
   \bibitem[{Kuraev(1976)}]{FKL2} 
   E.~A.~Kuraev, L.~N.~Lipatov and V.~S.~Fadin,
  {\it ``Multi - Reggeon Processes in the Yang-Mills Theory,''}
  Sov.\ Phys.\ JETP {\bf 44} (1976) 443
   [Zh.\ Eksp.\ Teor.\ Fiz.\  {\bf 71} (1976) 840].
   \bibitem[{Kuraev(1977)}]{FKL3}  
    E.~A.~Kuraev, L.~N.~Lipatov and V.~S.~Fadin,
 {\it ``The Pomeranchuk Singularity in Nonabelian Gauge Theories,''}
  Sov.\ Phys.\ JETP {\bf 45} (1977) 199
   [Zh.\ Eksp.\ Teor.\ Fiz.\  {\bf 72} (1977) 377].
   \bibitem[{Balitsky(1978)}]{BALI}
    I.~I.~Balitsky and L.~N.~Lipatov,
 {\it  ``The Pomeranchuk Singularity in Quantum Chromodynamics,''}
  Sov.\ J.\ Nucl.\ Phys.\  {\bf 28} (1978) 822
   [Yad.\ Fiz.\  {\bf 28} (1978) 1597].      
   \bibitem[{Lipatov(2017)}]{LIPEA}
     L.~N.~Lipatov,
 {\it  ``Effective actions for high energy scattering in QCD and in gravity,''}
  EPJ Web Conf.\  {\bf 164} (2017) 02002.
  doi:10.1051/epjconf/201716402002 and references therein.
  
  \bibitem[{Kovchegov(2012)}]{KOLEB}
  Y.~V.~Kovchegov and E.~Levin,
  {\it ``Quantum chromodynamics at high energy,''}
  Camb.\ Monogr.\ Part.\ Phys.\ Nucl.\ Phys.\ Cosmol.\  {\bf 33} (2012).
  \bibitem[{Ioffe(2010)}]{LIPB}
   B.~L.~Ioffe, V.~S.~Fadin and L.~N.~Lipatov,
  {\it ``Quantum chromodynamics: Perturbative and nonperturbative aspects,''}
  Camb.\ Monogr.\ Part.\ Phys.\ Nucl.\ Phys.\ Cosmol.\  {\bf 30} (2010).
  doi:10.1017/CBO9780511711817
  \bibitem[{Altinoluk(2010)}]{KLREG}
  T.~Altinoluk, A.~Kovner and E.~Levin,
  {\it ``Inside looking out: Probing JIMWLK wave functions with BFKL calculations,''}
  Phys.\ Rev.\ D {\bf 82} (2010) 074016
  doi:10.1103/PhysRevD.82.074016
  [arXiv:1004.5113 [hep-ph]].
  \bibitem[{Lipatov(1986)}]{LIPSOL}
  L.~N.~Lipatov,
  {\it ``The Bare Pomeron in Quantum Chromodynamics,''}
  Sov.\ Phys.\ JETP {\bf 63} (1986) 904
   [Zh.\ Eksp.\ Teor.\ Fiz.\  {\bf 90} (1986) 1536].
  
   \bibitem[{Ryskin(1980)}]{RYGL}
  M.~G.~Ryskin,
 {\it  ``High Transverse Momentum Hadron Production In The Pionization Region And Vacuum Singularity In QCD,'' }  Sov.\ J.\ Nucl.\ Phys.\  {\bf 32} (1980)  133,[ Yad.\ Fiz.\  {\bf 32} (1980) 259].
   \bibitem[{Levin(1980)}]{LERYDY}
  E.~M.~Levin and M.~G.~Ryskin,
  {\it ``Quantum Chromodynamics Prediction For Heavy Lepton Pair Production In Pionization Region.,''}  Sov.\ J.\ Nucl.\ Phys.\  {\bf 32} (1980) 413, [  Yad.\ Fiz.\  {\bf 32} (1980) 802].
   \bibitem[{Levin(1991)}]{LRSS}
    E.~M.~Levin, M.~G.~Ryskin, Y.~M.~Shabelski and A.~G.~Shuvaev,
  {\it ``Heavy quark production in semihard nucleon interactions,''}
  Sov.\ J.\ Nucl.\ Phys.\  {\bf 53} (1991) 657
   [Yad.\ Fiz.\  {\bf 53} (1991) 1059].
  \bibitem[{Gribov(1983)}]{GLR}
  L.~V.~Gribov, E.~M.~Levin and M.~G.~Ryskin,
 {\it  ``Semihard Processes in QCD,''}
  Phys.\ Rept.\  {\bf 100} (1983) 1.
  doi:10.1016/0370-1573(83)90022-4
      \bibitem[{Balitsky(1995)}]{BAL}
I.~Balitsky,
  {\it ``Operator expansion for high-energy scattering,''}
  Nucl.\ Phys.\ B {\bf 463}, 99 (1996)
  doi:10.1016/0550-3213(95)00638-9
  [hep-ph/9509348];\,\,\,
 {\it  ``Factorization and high-energy effective action,''}
  Phys.\ Rev.\ D {\bf 60} (1999) 014020,
[arXiv:hep-ph/9812311];\,\,\,\,

\bibitem[{Kovchegov(1999)}]{KOV}
Y.~V.~Kovchegov,
{\it ``Small x F(2) structure function of a nucleus including multiple pomeron exchanges,"}
 Phys.\ Rev. \, {\bf D60}, 034008  (1999),
[arXiv:hep-ph/9901281].

 
  \bibitem[{Mueller(1994)}]{MUCD}
 A.~H.~Mueller,
  {\it ``Soft Gluons In The Infinite Momentum Wave Function And The BFKL Pomeron,''}
  Nucl.\ Phys.\  B {\bf 415}, 373 (1994);
 {\it ``Unitarity and the BFKL pomeron,''}
  Nucl.\ Phys.\  B {\bf 437} (1995) 107
  [arXiv:hep-ph/9408245].
  \bibitem[{Mueller(2004)}]{MUSH}
    A.~H.~Mueller and A.~I.~Shoshi,
 {\it  ``Small x physics beyond the Kovchegov equation,''}
  Nucl.\ Phys.\ B {\bf 692} (2004) 175
  doi:10.1016/j.nuclphysb.2004.05.016
  [hep-ph/0402193].
  \bibitem[{Kovner(2016)}]{KLL}
  A.~Kovner, E.~Levin and M.~Lublinsky,
  JHEP {\bf 1608} (2016) 031
  doi:10.1007/JHEP08(2016)031
  [arXiv:1605.03251 [hep-ph]].
  \bibitem[{Levin(2014)}]{LILE}
   E.~Levin, L.~Lipatov and M.~Siddikov,
 {\it  ``BFKL Pomeron with massive gluons,''}
  Phys.\ Rev.\ D {\bf 89} (2014) no.7,  074002
  doi:10.1103/PhysRevD.89.074002
  [arXiv:1401.4671 [hep-ph]].
  \bibitem[{Levin(1990a)}]{LERY0}
  E.~M.~Levin and M.~G.~Ryskin,
  {\it ``The Shrinkage of the Diffraction Peak of the Bare Pomeron in {QCD},''}
  Sov.\ J.\ Nucl.\ Phys.\  {\bf 50} (1989) 881
   [Z.\ Phys.\ C {\bf 48} (1990) 231]
   [Yad.\ Fiz.\  {\bf 50} (1989) 1417].
  doi:10.1007/BF01554471
    \bibitem[{Levin(1990b)}]{LERY1}
   E.~M.~Levin and M.~G.~Ryskin,
 {\it  ``High-energy hadron collisions in QCD,''}
  Phys.\ Rept.\  {\bf 189} (1990) 267.
  doi:10.1016/0370-1573(90)90016-U
  
  \bibitem[{Levin(2013)}]{LETA}
  E.~Levin and S.~Tapia,
  {\it ``BFKL Pomeron: modeling confinement,''}
  JHEP {\bf 1307} (2013) 183
  doi:10.1007/JHEP07(2013)183
  [arXiv:1304.8022 [hep-ph]].
  
  \bibitem[{Levin(2015)}]{LEPION}
   E.~Levin,
 {\it  ``Large $\mathbf{b}$ behaviour in the CGC/saturation approach: BFKL equation with pion loops,''}
  Phys.\ Rev.\ D {\bf 91} (2015) no.5,  054007
  [arXiv:1412.0893 [hep-ph]].
  \bibitem[{Kovner(2002)}]{KW}
    A.~Kovner and U.~A.~Wiedemann,
 {\it  ``Nonlinear QCD evolution: Saturation without unitarization,''}
  Phys.\ Rev.\ D {\bf 66}, 051502 (2002)
  [hep-ph/0112140];\,\,
  {\it ``Perturbative saturation and the soft pomeron,''}
  Phys.\ Rev.\ D {\bf 66}, 034031 (2002)
  [hep-ph/0204277];\,\,
  {\it ``No Froissart bound from gluon saturation,''}
  Phys.\ Lett.\ B {\bf 551}, 311 (2003)
  [hep-ph/0207335].
  \bibitem[{Lipatov(1976)}]{LIPHO1}
     L.~N.~Lipatov,
  {\it ``Calculation of the Gell-Mann-Low Function in a Scalar Field Theory with Strong Nonlinearity,''}
  Sov.\ Phys.\ JETP {\bf 44} (1976) 1055
   [Zh.\ Eksp.\ Teor.\ Fiz.\  {\bf 71} (1976) 2010].
     \bibitem[{Lipatov(1977a)}]{LIPHO2}   
     L.~N.~Lipatov,
   {\it ``Divergence of the Perturbation Theory Series and the Quasi-classical Theory,''}
  Sov.\ Phys.\ JETP {\bf 45} (1977) 216
   [Zh.\ Eksp.\ Teor.\ Fiz.\  {\bf 72} (1977) 411]. 
        \bibitem[{Lipatov(1977b)}]{LIPHO3}     
 L.~N.~Lipatov,
   {\it ``Divergence of Perturbation Series and Pseudoparticles,''}
  JETP Lett.\  {\bf 25} (1977) 104
   [Pisma Zh.\ Eksp.\ Teor.\ Fiz.\  {\bf 25} (1977) 116]. 
    \bibitem[{Lipatov(1979)}]{LIPHO4}   
   L.~N.~Lipatov, A.~P.~Bukhvostov and E.~I.~Malkov,
{\it ``Large Order Estimates for Perturbation Theory of a Yang-Mills Field Coupled to a Scalar Field,''}
  Phys.\ Rev.\ D {\bf 19} (1979) 2974.
  \bibitem[{Bukhvostov(1977)}]{LIPHO5}
      A.~P.~Bukhvostov and L.~N.~Lipatov,
 {\it ``High Orders of the Perturbation Theory in Scalar Electrodynamics,''}
  Phys.\ Lett.\  {\bf 70B} (1977) 48.
  \bibitem[{Fadin(1998)}]{FALINLO}
   V.~S.~Fadin and L.~N.~Lipatov,
  {\it ``BFKL pomeron in the next-to-leading approximation,''}
  Phys.\ Lett.\ B {\bf 429} (1998) 127
  doi:10.1016/S0370-2693(98)00473-0
  [hep-ph/9802290].
  \bibitem[{Salam(1998)}]{SAL}
  G.~P.~Salam,
 {\it  ``A Resummation of large subleading corrections at small x,''}
  JHEP {\bf 9807} (1998) 019
  doi:10.1088/1126-6708/1998/07/019
  [hep-ph/9806482].
  \bibitem[{Ciafaloni(1999)}]{CCS}
   M.~Ciafaloni, D.~Colferai and G.~P.~Salam,
 {\it  ``Renormalization group improved small x equation,''}
  Phys.\ Rev.\ D {\bf 60} (1999) 114036
  doi:10.1103/PhysRevD.60.114036
  [hep-ph/9905566].
  \end{thebibliography}
 \end{document}